\documentstyle[12pt,aasms4]{article}

\received{1998 September 22}
\accepted{1998 November 11}

\slugcomment{To appear in {\it The Astrophysical Journal}}

\lefthead{Crenshaw, et al.}
\righthead{Intrinsic Absorption in Seyfert 1 Galaxies}

\begin{document}

\title{Intrinsic Absorption Lines in Seyfert 1 Galaxies.\\
I. Ultraviolet Spectra from the Hubble Space Telescope\altaffilmark{1}}

\author{D. Michael Crenshaw\altaffilmark{2,3},
Steven B. Kraemer\altaffilmark{2},
Albert Boggess\altaffilmark{4,5},\\
Stephen P. Maran\altaffilmark{5,6}, 
Richard F. Mushotzky\altaffilmark{7},
and Chi-Chao Wu\altaffilmark{8}}

\altaffiltext{1}{Based on observations made with the NASA/ESA Hubble Space 
Telescope, obtained from the data archive at the Space Telescope Science 
Institute. STScI is operated by the Association of Universities for Research in 
Astronomy, Inc. under the NASA contract NAS5-26555. }

\altaffiltext{2}{Catholic University of America and
Laboratory for Astronomy and Solar Physics,
NASA's Goddard Space Flight Center, Code 681
Greenbelt, MD  20771.}

\altaffiltext{3}{Email: crenshaw@buckeye.gsfc.nasa.gov.}

\altaffiltext{4}{2420 Balsam Drive, Boulder, CO  80304.}

\altaffiltext{5}{NASA's Goddard High Resolution Spectrograph (GHRS) 
Investigation Definition Team.}

\altaffiltext{6}{Space Sciences Directorate, Code 600,
NASA's Goddard Space Flight Center, Greenbelt, MD  20771.}

\altaffiltext{7}{Laboratory for High Energy Astrophysics,
Code 662, NASA's Goddard Space Flight Center, Greenbelt, MD  20771.}

\altaffiltext{8}{Computer Sciences Corporation,
Space Telescope Science Institute, 3700 San Martin Drive,
Baltimore, MD  21218.}

\begin{abstract}

We present a study of the intrinsic absorption lines in the ultraviolet spectra 
of Seyfert 1 galaxies. The study is based on spectra from the {\it Hubble Space 
Telescope}, and includes the Seyfert 1 galaxies observed with the Faint 
Object Spectrograph and Goddard High Resolution Spectrograph at spectral 
resolutions of $\lambda$/$\Delta\lambda$ $\approx$ 1000 -- 20,000 with 
good signal-to-noise ratios. We find that the fraction of Seyfert 1 
galaxies that show intrinsic absorption associated with their active nuclei is 
more than one-half (10/17), which is much higher than previous 
estimates (3 -- 10\%) based on {\it IUE} data. There is a one-to-one 
correspondence between Seyferts that show intrinsic UV absorption and X-ray 
``warm absorbers'', indicating that these two phenomena are 
related. Although our sample is not complete, we 
conclude that intrinsic absorption represents an important component that needs 
to be integrated into our overall physical picture of active galaxies.

The intrinsic UV absorption is generally characterized by high-ionization: C~IV 
and N~V are seen in all 10 Seyferts with detected absorption (in addition to 
L$\alpha$), whereas
Si~IV is present in only four of these Seyferts, and Mg~II absorption is only 
detected in NGC~4151. The  absorption lines are blueshifted (or in a few 
cases at rest) with respect to the 
narrow emission lines, indicating that the absorbing gas is undergoing net 
radial outflow. At high resolution, the absorption often splits into distinct 
kinematic components that show a wide range in widths (20 -- 400 km s$^{-1}$ 
FWHM), indicating macroscopic motions (e.g., radial velocity subcomponents or 
turbulence) within a component. The strong absorption components have cores 
that are much deeper than the continuum flux levels, indicating that the regions 
responsible for these components lie completely outside of the broad 
emission-line regions. 

Additional information on the covering factors and column densities can be 
derived from the absorption profiles in the high resolution spectra.
The covering factor of the absorbing gas in the line of sight, relative to the 
total underlying emission, is C$_{los}$ $\geq$ 0.86, on average.
The global covering factor, which is the fraction of emission intercepted by the 
absorber averaged over all lines of sight, is C$_{global}$ $\geq$ 0.5.
Thus, structures covering large solid angles as seen by the central 
continuum source (e.g., spherical shells, sheets, or cones with large opening 
angles) are required. The individual absorption components show a wide range in 
C~IV column densities 
(0.1 -- 14 x 10$^{14}$ cm$^{-2}$), and the ratio of N~V to C~IV column density 
varies significantly from one absorption component to the next, even in the same 
Seyfert galaxy. Thus, the intrinsic absorption in a Seyfert 1 galaxy is 
typically comprised of distinct kinematic components that are characterized by 
a range in physical conditions (e.g., ionization parameter and hydrogen 
column density).

Finally, we show evidence for extreme variability in the intrinsic absorption 
lines of NGC 3783. In addition to our earlier report of the appearance of a C~IV 
absorption doublet at $-560$ km s$^{-1}$ (relative to the emission lines) 
over 11 months, we have detected the appearance of another C~IV doublet at 
$-1420$ km s$^{-1}$ over 15 months. On the other hand, the C~IV absorption lines 
of NGC 3516 and NGC 4151 were very stable over periods of 6 months and 4 years, 
respectively. Monitoring observations of individual Seyferts at higher 
time resolution are needed to distinguish between different sources of 
variability (variable ionization, motion of gas across the line of sight) and 
to determine the densities and radial locations of the absorption components.

\end{abstract}


\section{Introduction}

Seyfert galaxies were originally recognized as a special class of objects from 
their strong emission lines (Seyfert 1943), and were later subdivided into two 
groups based on the widths of their emission lines in optical spectra 
(Khachikian \& Weedman 1971). 
Seyfert 1 galaxies have broad permitted lines with widths $\geq$ 1000 km 
s$^{-1}$ (FWHM), and narrow permitted and forbidden lines with widths $\approx$ 
500~km~s$^{-1}$ (FWHM), whereas Seyfert 2 galaxies show only the narrow 
emission lines.
In turn, these objects are recognized as belonging to the category of 
objects known as active galaxies (Osterbrock 1984). It has been known for some 
time that Seyfert galaxies are strong 
emitters over the entire electromagnetic spectrum (Weedman 1977). The optical 
continua of Seyfert 1 galaxies are dominated by nonstellar emission that can be 
characterized by a power-law, as first demonstrated by Oke \& Sargent (1968), 
whereas this component appears to be much weaker in Seyfert 2 galaxies (Koski 
1978). These trends extend to the lines and continua observed in the 
ultraviolet 
(Wu, Boggess, \& Gull 1983) by the {\it International Ultraviolet Explorer} 
({\it IUE}). 

Seyfert (1943) recognized the presence of absorption lines in several of his 
objects, which he attributed to a ``G-type spectrum''. In fact, most of the 
absorption features in the {\it optical} spectra of both types of Seyfert can 
be 
attributed to stellar features from the host galaxy (Koski 1978; Crenshaw \& 
Peterson 1985). Oke \& Sargent (1968) first reported a possible nonstellar 
absorption feature in the spectrum of NGC 4151; they explain that this feature 
was first discovered and attributed to He I $\lambda$3889 self-absorption by 
O.C. Wilson. Anderson \& Kraft (1969) resolved the metastable He I absorption 
into three components, and discovered hydrogen Balmer (H$\beta$, H$\gamma$) 
absorption corresponding to these components. The components are all 
blue-shifted relative to the emission lines by up to $-$970 km s$^{-1}$, and 
the authors attributed the blue-shifts to ejection of matter from the nucleus of 
NGC 4151. Cromwell and Weymann (1970) showed that the strengths of the Balmer 
absorption lines are variable, and can occasionally disappear altogether. 
Anderson (1974) found evidence that the Balmer absorption lines in NGC 4151 can 
vary on time scales as small as $\sim$30 days, and claimed that the absorption 
and continuum variations 
are correlated. Anderson also suggested that the absorption region lies outside 
of the broad emission-line region, because the depth of the H$\alpha$ 
absorption was greater than the flux in the nearby continuum.

In the ultraviolet, early observations of NGC 4151, primarily by {\it IUE}, 
revealed a rich assortment of intrinsic absorption lines spanning a wide range 
in ionization from Mg~II to N~V, as well as fine-structure and metastable 
absorption lines such as C III$^{*}$ $\lambda$1175 (Davidsen \& Hartig 1978; 
Boksenberg et al. 1978; Penston et al. 1981; Bromage et al. 1985). Bromage et 
al. suggested that the absorption-line variations in NGC 4151 are due to changes 
in the column densities, rather than the velocity spread of the clouds. In 
addition, they suggested that the absorbing region is likely to be composed of 
optically thin gas located outside of the broad emission-line region.
Ulrich (1988) described other Seyfert galaxies that 
show obvious intrinsic absorption in {\it IUE} spectra. Most of these objects 
show C~IV and N~V absorption (along with L$\alpha$ in most cases), indicating 
high ionization. Only two Seyferts showed strong evidence for intrinsic Mg II 
absorption (NGC 4151 and MCG 8-11-11). The absorption features are 
blue-shifted by as much as $-$2500~km~s$^{-1}$ with respect to the emission 
lines.

From a large set of {\it IUE} observations, Ulrich (1988) estimated that only 3 
-- 10\% of all Seyfert 1 galaxies show intrinsic UV absorption. This would
suggest that while intrinsic UV absorption is an interesting curiosity, it 
might not have a strong impact on our understanding of active galaxies. 
However, only very strong absorption lines (with equivalent widths~$>$~1~\AA) 
are detectable at the low spectral resolution ($\lambda$/$\Delta\lambda$ $=$ 
200 
-- 400) and low signal-to-noise ratio (SNR $<$ 10) of the {\it IUE} spectra. 
From an examination of a small sample of {\it HST} spectra, we found that more 
than half (5/8) of the Seyfert 1 galaxies showed intrinsic C~IV absorption 
(Crenshaw 1997). This indication that intrinsic UV absorption is much more 
common than previously believed was a principal motivation for expanding our 
sample of {\it HST} spectra, as described in this paper.

Another motivation for this study was to investigate the relationship between 
the UV absorber and the X-ray ``warm absorber'', which is characterized by 
O~VII and O~VIII absorption edges and is present in about half of the Seyfert 1 
galaxies observed in X-rays (Reynolds 1997; George et al. 1998b). Evidence 
for this highly ionized gas in the line-of-sight to an active galaxy was first
reported by Halpern (1984), using the {\it Einstein Observatory}. Subsequent 
X-ray satellites provided supporting evidence for warm absorbers, and detailed 
analysis and modeling of many individual objects were made possible by {\it 
ASCA}. Warm absorbers are typically 
characterized by ionization parameters in the range U $=$ 0.1 -- 10, 
temperatures on the order of 10$^{5}$ K, and ionized hydrogen column densities 
in the range 10$^{21}$ -- 10$^{23}$ cm$^{-2}$ (Reynolds \& Fabian 1995; George 
et al. 1998b).

A connection between the UV and X-ray absorbers was made by Mathur et al. 
(1994), based on quasi-simultaneous {\it ROSAT} and {\it HST} observations of 
the quasar 3C~351. Using photoionization models, these authors found that a 
single component of ionized gas could produce both the observed strengths of 
the O~VII and O~VIII absorption edges and the equivalent widths of the 
intrinsic UV absorption lines (O~VI $\lambda\lambda$1031.9, 1037.6; N~V 
$\lambda\lambda$1238.8, 1242.8; C~IV $\lambda\lambda$1548.2, 1550.8).
In these models, most of the carbon, nitrogen, and oxygen are in higher 
ionization states than those represented by the UV lines.
These authors claim that the UV and X-ray absorbers are also the same in 3C~212 
(Mathur 1994) and the Seyfert 1 galaxy NGC 5548 (Mathur et al. 1995).
However, in some objects, it appears that multiple components characterized by 
a wide range in ionization paramater and hydrogen column density are needed to 
explain the wide range in ionization species; such is the case for NGC 3516 
(Kriss et al. 1996a; Crenshaw et al. 1998) and NGC 4151 (Kriss et al. 1995).
To further investigate the proposition that a single component is responsible 
for both the UV and X-ray absorption, we felt that it was important to study the 
absorption in other sources, particularly those with both {\it HST} and 
{\it ASCA} spectra. 

Finally, we were motivated by the fact that the intrinsic UV absorption in 
Seyfert galaxies is qualitatively similar to the ``associated'' absorption seen 
in QSOs, in that the lines are relatively narrow ($\leq$~300~km~s$^{-1}$), 
usually close to the emission-line redshift, 
and in some cases, variable (Hamann 1997; Hamann et al. 1997). The relationship 
between these relatively narrow absorption systems and the broad absorption 
lines (BALS), which can exhibit 
troughs reaching blueshifts as high as $\sim$0.1c, and are found in 
approximately 10\% of all QSOs (Weymann et al. 1981; Weymann et al. 1991), is 
not known. The fact that these systems represent high-ionization gas flowing 
outward from the nucleus suggests that these phenomena may be related. Thus, 
detailed studies of the intrinsic absorption in Seyfert 1 galaxies should lead 
to a better understanding of outflows in active galaxies as a function of 
luminosity. Eventually, we hope that these studies will provide clues to the 
physical structure of active galaxies, and the relationship between the 
various absorption, emission, and continuum components in these objects.

To gain a better understanding of the intrinsic absorption in Seyfert 1 
galaxies, we have obtained all of the UV spectra of these objects from 
the {\it HST} data archive. In this paper, we present new results on the basic 
properties of the intrinsic UV absorber, its connection to the X-ray warm 
absorber, its frequency of occurrence in Seyfert 1 galaxies, its global 
covering factor (relative to the continuum source and broad emission-line 
region), and its variability. We give a brief history of previous ultraviolet 
observations of each object in Appendix A, and discuss a few interesting objects 
that did not satisfy our selection criteria in Appendix B.

\section{Observations and Data Reduction}

Our original sample was a collection of eight Seyfert 1 galaxies observed under 
the GHRS GTO (Guaranteed Time Observations) programs of Boggess and Maran; 
preliminary results on detection of intrinsic absorption were reported in 
Crenshaw (1997). Results on a few observations from our sample were 
previously reported for Mrk 509 (Crenshaw, Boggess, \& Wu 1993), NGC 3783 (Maran 
et al. 1996), and NGC 3516 (Crenshaw, Maran, \& Mushotzky 1998).

To increase the sample, we searched the {\it HST} Data Archive to find all of 
the objects that were 
identified as Seyfert galaxies by the original proposers and have ultraviolet 
spectra obtained by the Faint Object Spectrograph (FOS) or the Goddard High 
Resolution Spectrograph (GHRS). Our search was conducted one year after 
the removal of these two instruments from {\it HST} (in 1997 February), so
all of the data were nonproprietary and available for retrieval. We did not 
include the few spectra obtained with the Faint Object Camera (FOC), as they do 
not satisfy the criteria given below. We included only those objects in this 
group that have been identified 
as Seyfert 1, Seyfert 1.5, or narrow-line Seyfert 1 galaxies (Osterbrock \& 
Pogge 1985) in the literature or in the NASA/IPAC Extragalactic Database (NED);
hereafter, we refer to this group in a general way as Seyfert 1 galaxies. We 
excluded Seyfert 2 galaxies or those objects identified as Seyfert 1.8 or 1.9 
galaxies (cf., Osterbrock 1981). The reason for excluding these objects is a 
practical one: it is difficult to detect the absorption lines if they are not 
seen against strong and relatively broad emission lines and/or strong UV 
continua. We identified $\sim$40 Seyfert 1 galaxies with UV spectra in the 
archives. 

After retrieving and inspecting the UV spectra, we used the following additional 
criteria for including objects in our survey:\\
1) There must be at least one spectral region in which an intrinsic absorption 
line might be detected. In practice, we included objects that had an observation
of the redshifted N~V, Si~IV, C~IV, or Mg~II emission-line profiles,
since the strongest absorption lines come from these species and are embedded in 
these profiles (Ulrich 1988). We did not include objects that were only observed 
in the L$\alpha$ region, since this absorption line can originate in the 
low-redshift L$\alpha$ forest (cf., Stocke et al. 1995; Shull et al. 1996).\\
2) The resolution must be sufficient to detect and, in most cases, separate the 
absorption doublets. In practice, we required that $\lambda$/$\Delta\lambda$ 
must be $\geq$ 1000, where $\lambda$ is the observed central wavelength and 
$\Delta\lambda$ is a resolution element. Henceforth, we refer to
FOS spectra with $\lambda$/$\Delta\lambda$ $\approx$ 1000 and GHRS spectra 
with  $\lambda$/$\Delta\lambda$ $\approx$ 2000 as the ``low resolution 
spectra'', and GHRS spectra with $\lambda$/$\Delta\lambda$ $\approx$ 20,000
as the ``high resolution spectra''.\\
3) The signal-to-noise ratio (SNR) per resolution element near the 
expected location of absorption must be $\geq$ 20 for the low resolution 
spectra, and $\geq$ 8 for the high resolution spectra. The detection limit for 
absorption lines varies according to the SNR; a typical 3$\sigma$ 
detection limit (in equivalent width) is $\sim$0.1 \AA\ for the low resolution 
spectra, and $\sim$0.03 \AA\ for the high resolution spectra.

We found that 17 Seyfert 1 galaxies in the {\it HST} data archive satisfied the 
above criteria. Only a few of the rejected spectra did not satisfy criteria 1 
or 2; most of them were rejected on the basis of poor SNR. We 
emphasize that the observations in our sample were acquired by various 
investigators for a variety of purposes, and that we did not include QSOs at
low redshifts. Thus, our sample is obviously {\it not} a complete one.
In Appendix B, we list those Seyferts that did not satisfy our 
criteria but show possible evidence for intrinsic absorption.

Table 1 lists the low and high resolution observations for the 17 Seyfert 1 
galaxies in our sample. Note that each object has coverage of the redshifted 
L$\alpha$, N~V, Si~IV, and C~IV lines, whereas 15 objects have coverage of the 
redshifted Mg~II region (NGC 3783 and II~Zw~136 do not have coverage of this 
region). Spectra that were taken within days of each 
other have been averaged to increase the SNR. There are multiple epochs of 
observation for a number of these objects, 
although each observation does not necessarily contain every wavelength 
interval. High resolution observations are available for five 
Seyfert 1 galaxies observed at low resolution. For these observations, we list 
the redshifted emission-line 
profiles that are present in each wavelength setting, where we expect 
to find the intrinsic absorption lines.

We used the following data reduction procedures for the spectra in Table 1.
For the FOS spectra, we retrieved the calibrated files from the {\it HST} data 
archive to obtain the wavelength, absolute flux, error flag, and photon noise 
vectors for each spectrum. For the GHRS spectra, we reduced the raw data
using the IDL procedures written for the GHRS Instrument 
Definition Team (Blackwell et al. 1993); in the process, we determined an 
average background across the diode array and subtracted it from the gross 
spectrum for each readout, as described in Crenshaw et al. (1998).
We interpolated all of the spectra to a linear wavelength scale, retaining the 
original approximate wavelength intervals (in \AA\ per bin), and, for each 
object, we averaged all of the spectra obtained at a particular wavelength 
setting and within a few days of each other. We then combined spectra that were 
obtained at different wavelength settings and observed within days 
of each other at a single point in the wavelength region 
of overlap. In the few instances where the fluxes did not agree to within 5\% 
in the region of overlap, the longer wavelength spectrum was scaled in flux to 
match the shorter wavelength spectrum. Thus, we have a final UV spectrum for 
each entry in Table 1. 

In each spectrum, we identified lines that arise in the interstellar 
medium or halo of our Galaxy using the lists of UV resonance lines in Morton, 
York, \& Jenkins (1988).  Those lines that could not be identified as Galactic 
in origin were considered to be candidates for intrinsic absorption. 
Specifically, absorption lines with heliocentric velocities greater than a few 
hundred km s$^{-1}$, at the proper wavelength separations for the doublets, and 
with equivalent widths $\geq$ 3 times the measurement errors (see below) were 
considered to be intrinsic. In addition, only L$\alpha$ lines that occured at 
radial velocites close to those of other detected lines were considered to be 
intrinsic, since those at other velocities could be due to the low-redshift 
L$\alpha$ forest.

Figures 1 and 2 present the far-UV (1200 -- 1700 \AA) portion of the 
low resolution spectra. The locations of the strongest Galactic 
absorption lines (Morton et al. 1988) are also shown. Figure 1 shows the seven 
Seyfert 1 galaxies that did not show detectable intrinsic absorption.  Figure 2 
shows the ten Seyfert 1 galaxies with intrinsic  
L$\alpha$, N~V~$\lambda\lambda$1238.8, 1242.8,
Si~IV~$\lambda\lambda$1393.8, 1402.9,
and/or C~IV~$\lambda\lambda$1548.2, 1550.8
absorption lines . Note that the intrinsic absorption lines span a 
large range in equivalent width and velocity width, and tend to occur in the 
cores of the emission-line profiles or are blueshifted with respect to the 
emission lines. In a few cases, multiple velocity components can be seen, even 
at this relatively low resolution.

Figures 3 and 4 give the high-resolution spectra of Seyfert 1 galaxies with 
intrinsic absorption. Figure 3 shows the C~IV region in four Seyferts, and 
demonstrates the necessity of high spectral resolution for resolving absorption 
components. Figure 4 shows the corresponding L$\alpha$ and N~V components, when 
available.

\section{Direct measurements}

We used the following procedures to measure the intrinsic absorption lines 
that we detected. To determine the shape of the underlying emission, we fit a 
cubic spline to regions on either side of the absorption. For those cases in 
which the absorption occurs in the very core of the 
line, this procedure may underestimate the underlying emission if there is a 
significant narrow-line contribution that has not been properly taken into 
account. Based on 
visual inspection of the unabsorbed profiles in the same spectrum (e.g., C~III] 
$\lambda$1909), we do not expect this to be a serious problem. To normalize the 
absorption profiles, we divided the observed spectra by the cubic spline fits. 
We then made direct measurements of the centroid, equivalent width 
(EW), and full-width at half-maximum (FWHM) of each visually distinct 
absorption component. Obviously, the number of visible components may depend on 
spectral resolution, so we consider the low and high resolution spectra 
separately. We note that our distinction between components and structure within 
components is somewhat subjective.

We determined the uncertainites in the centroids and FWHMs of each feature from 
different reasonable fits to the emission on either side of the absorption 
lines. We determined uncertainties in EWs from the sum in quadrature of the 
error in fitting the underlying emission and the uncertainty due to photon 
noise. Finally, we excluded from further consideration any absorption feature 
with an EW less than three times its uncertainty.

\subsection{Detection of intrinsic absorption}

Table 2 provides information on the redshifts of Seyfert 1 galaxies in our 
sample; 4 of these 17 objects are generally regarded as 
narrow-line Seyfert 1 galaxies, as noted. Redshifts 
from H~I observations (when available) and from the narrow optical 
emission lines (primarily [O~III] $\lambda\lambda$4959, 5007) were obtained 
from NED. The H~I redshifts are probably 
more indicative of the systemic velocities of the host galaxies, but we have 
values for only eight galaxies. The optical redshifts tend to be slightly lower, 
on average, but within $\sim$100 km s$^{-1}$ of the neutral hydrogen redshifts. 
To be consistent, we adopt the optical redshifts when determining the radial 
velocities of the absorption lines relative to the Seyfert galaxy, with the 
understanding that the absorption velocities relative to the systemic redshift 
are likely to be within 100 km s$^{-1}$ of these values. We note that in a 
few cases, the absorption lines in Figure 2 appear to be slightly 
redshifted with respect to the cores of the broad emission lines (e.g., in Mrk 
509), due to a slight blueshift of the high ionization emission lines with 
respect to the lower ionization lines (see Gaskell 1982).

In Table 2, we specify whether or not each Seyfert 1 galaxy shows
intrinsic UV absorption, based on our criteria and according to the detection 
limits that we gave in the previous section. The 
Seyferts with intrinsic C~IV absorption always show N~V and L$\alpha$ absorption 
at the same approximate velocities, as we discuss in more detail later.
We also specify whether or not each Seyfert 1 galaxy shows evidence for an 
X-ray warm absorber (Reynolds 1997; George et al. 1998b). This table confirms
our earlier finding that intrinsic UV absorption is very common: more than half 
(10/17) of the Seyfert 1 galaxies show intrinsic absorption lines.

\subsection{Results from the direct measurements}

Tables 3 and 4 present direct measurements of the intrinsic absorption lines in 
the low and high resolution spectra, respectively. 
We give the observed centroid ($\lambda_{obs}$), the equivalent width (EW), our 
identification, and the radial velocity, relative to the optical redshift, for 
each absorption feature.  Where there is evidence for more than one kinematic 
component, we number each component beginning with the one at the shortest 
observed wavelength. 
In a few instances, an intrinsic absorption component is blended with another 
Galactic or intrinsic line; in these cases, we identify the contributors and 
give the centroid for the entire feature, but do not give other direct 
measurements.
For the high resolution spectra in Table 4, we also give the full-width at 
half-maximum (FWHM) of each component, when possible, since the absorption 
features are resolved (the instrumental FWHM is $\sim$15 km s$^{-1}$).
Some components are blended enough to prevent an accurate measure of their 
widths.
 
Measurement errors for $\lambda_{obs}$ are typically 0.05 -- 0.1 \AA\ (10 -- 20 
km s$^{-1}$) for the low resolution spectra and 0.02 \AA\ ($\sim$4 km s$^{-1}$) 
for the high resolution spectra; measurement errors for FWHM are typically 
$\sim$10 km s$^{-1}$. We provide more realistic uncertainties in v$_{r}$ and 
FWHM below, by comparing values for all of the ions associated with each 
kinematic component. 

We summarize the kinematic components that we have detected in Tables 5 and 6.
The radial velocity centroids and FWHM are averages from all of the
lines that we think are associated with one kinematic component, and the
uncertainties are standard deviations of the averages. As we noted earlier, the 
number of components is highly dependent on the spectral resolution. For 
objects 
that have both high  and low resolution spectra, it is clear that the 
low resolution values are just ``averages'' for blends of the strongest lines. 
In Mrk 509, there is an offset in v$_{r}$ between the low and high resolution 
spectra of $\sim$80 kms$^{-1}$, which is probably due to a zero-point error in 
the low resolution wavelength scale of $\sim$0.3~\AA. Similar errors in v$_{r}$ 
are possible for the other low resolution spectra.

\subsection{Conclusions based on direct measurements}

We can draw some important conclusions from Tables 2, 5, and 6 and Figures 2 -- 
4, to support and extend previous results on the intrinsic UV absorption lines 
(which were based on a few Seyfert 1 galaxies, see section 1):

1. From the low resolution sample, the frequency of occurence of intrinsic 
absorption is f $=$ 10/17 $=$ 0.59. When intrinsic absorption is detected, high 
ionization lines (C~IV, N~V) and L$\alpha$ are always seen. The intermediate 
ionization line of Si~IV is seen in only 4/17 of the Seyfert 1 galaxies. Of the 
15 objects with near-UV coverage, only NGC 4151 shows evidence for 
low-ionization (Mg~II) absorption,

2. The centroids of the absorption lines are blueshifted by up to $-$ 2100 km  
s$^{-1}$, or occasionally at rest, with respect to the narrow emission
lines (and relative to neutral hydrogen, when it is detected). 
Thus, the UV absorbers are, for the most part, undergoing radial outflow 
with respect to the nuclei of these Seyfert 1 galaxies.

3. At high resolution, the absorption lines often split into distinct kinematic
components. The components are resolved, and exhibit a range of widths (20 --
400 km~s$^{-1}$~FWHM). Most of the components have widths much larger than that 
expected for thermally stable photoionized gas, even if they arise in the warm 
absorber (FWHM of C~IV $\approx$ 20 km s$^{-1}$ at 10$^{5}$ K). This indicates 
macroscopic motions within a component, such as turbulence or the superposition 
of additional radial velocity components.

4. Lines that are narrow and within a few hundred km s$^{-1}$ of the systemic 
radial velocity could potentially come from the interstellar medium or halo of 
the host galaxy. However, most of the intrinsic absorption lines arise close to 
the nucleus, because they are broad, have large blueshifts, and/or they are 
variable (see section 5).

5. In the high resolution spectra, the cores of the strongest absorption 
components are deeper than the heights (in flux) of the continua. This indicates 
that both the central continuum source and at least a portion of the BLR is 
occulted by the UV absorber. When we know the size of the BLR (e.g., from 
reverberation mapping experiments), we can determine lower limits to the size of 
the absorber in the plane of the sky and its distance from the continuum source 
(see section 4.3).

6. From Table 3, we can see that there is a one-to-one correspondence between 
those Seyferts that show UV and X-ray absorption in this sample; six objects 
show both UV and 
X-ray absorption, and two objects show neither (although the case for a 
warm absorber in NGC 7469 is not strong). Thus, although the 
sample is small, it appears that there is a connection between the UV and X-ray 
warm absorbers; however, this does not require that they arise from the same 
zone, characterized by a single ionization parameter.

\section{Absorption profiles}

Additional information can be derived from the resolved absorption lines in the 
high resolution spectra.
We have omitted the low resolution spectra from this analysis for the following 
reasons.
1) The absorption lines seen at low resolution are often comprised of
multiple kinematic components, and these components are likely to be 
characterized by different physical conditions.
2) It is more difficult to determine the intrinsic shapes of the underlying 
profiles in the low resolution spectra, since the absorption components are 
broadened and blended.
3) Many of the absorption lines are saturated in the low resolution 
spectra, which typically leads to underestimates of the column densities
\footnote{Lower limits to the column densities of the absorption lines in the 
low resolution spectra can be calculated directly from the EWs in Table 3, by 
assuming the case of unsaturated lines (Spitzer 1978). Other techniques for 
estimating the column densities from unresolved or marginally resolved 
absorption lines include the curve-of-growth method (Spitzer 1978), and the 
apparent optical depth method (Savage \& Sembach 1991).  
These techniques are based on simplifying assumptions, which are discussed in 
these references.}.

\subsection{Covering factors}

The methods we used to analyze the absorption profiles follow those of 
Hamann et al. (1997). We consider the 
case where the region responsible for an absorption component does 
not completely cover the emission source(s) behind it. If this effect is present 
and not corrected for, the column densities will be underestimated. 
We define the covering factor, C$_{los}$, to be the fraction of continuum plus 
BLR that is occulted by the absorber in our line of sight. We will 
consider the case where the covering factor could differ from one kinematic 
component to the next (but not within a component, as discussed below). If 
we normalize the absorption lines by dividing by the underlying emission, and 
I$_r$ is the residual flux in the core of the absorption line at a particular 
radial velocity, then we can determine a lower limit to the covering factor in 
the line of sight from the residual flux:
\begin{equation}
C_{los} \geq 1 - I_r.
\end{equation}

We can also determine a lower limit to the fraction of BLR flux that is occulted 
by the absorber:
\begin{equation}
C_{los}^{BLR} \geq \frac{1 - I_c - I_r}{1 - I_c},
\end{equation}

where I$_c$ is the fraction of the underlying emission at a particular radial 
velocity due to the central continuum source.

The actual value of C$_{los}$ for a component can be determined from a doublet, 
if both lines of the doublet are unblended with other lines (Hamann et al. 
1997). If the expected ratio of the 
optical depths of the doublet is 2 (as for C~IV and N~V), then
\begin{equation}
C_{los} = \frac{I_1^2 - 2I_1 + 1}{I_2 - 2I_1 + 1},
\end{equation}

where I$_1$ and I$_2$ are the residual fluxes in the cores of the weaker line 
(e.g., C~IV $\lambda$1550.8) and stronger line (e.g., C~IV $\lambda$1548.2), 
respectively.

A lower limit to C$_{los}$ is easily obtained, particularly if the 
absorption components are resolved. To determine a well-constrained value of  
C$_{los}$, however, we need both members of the doublet to be clean (unblended 
with other lines), reasonably strong lines, high SNR, and an 
accurate estimate of the underlying emission. In addition, since even the strong 
components are affected by blending with other components in their wings, we 
are constrained to determining C$_{los}$ in the cores of these components.

In practice, we determined the covering factor for a few (2 -- 5) 
resolution elements across the core of each component, and then calculated the 
average and uncertainty.
Actual values of the covering factors, as opposed to lower limits, were 
determined only for a few strong, unblended components (see Figures 3 and 4).
For the lower limits, we chose those lines that yield the highest 
values for that component (implicitly assuming that for a given kinematic 
component, the covering factor is the same for L$\alpha$, C~IV, and N~V).

Table 6, gives C$_{los}$ and C$_{los}^{BLR}$ 
for each kinematic component in the high resolution spectra, along with the 
line used to determine these values. We note that when actual values are 
determined, they are close to unity (except for NGC 3783).

\subsection{Column densities}

Since the absorption components are resolved, we can determine the column 
density of each component from its optical depth as a function of radial 
velocity (v$_r$). If the covering factor C$_{los}$ $=$ 1 for a component, then 
the optical depth at a particular v$_r$ is given by:
\begin{equation}
\tau = ln \left(\frac{1}{I_{r}}\right),
\end{equation}

If C$_{los}$ $\neq$ 1, then the optical depth at a particular radial velocity 
is:
\begin{equation}
\tau = ln \left(\frac{C_{los}}{I_r + C_{los} -1}\right) 
\end{equation}
(Hamann et al. 1997). In principal, C$_{los}$ can be a function of radial 
velocity (v$_r$), but we have assumed that it is constant for a given component.

The column density is then obtained by integrating the optical depth across the 
profile:
\begin{equation}
N = \frac{m_{e}c}{\pi{e^2}f\lambda}~\int \tau(v_{r}) dv_{r}
\end{equation}
(Savage \& Sembach 1991), where f is the oscillator strength and $\lambda$ is 
the laboratory wavelength (Morton et al. 1988).

In Table 7, we give the column densities for each component in the 
high resolution spectra, for the case of C$_{los}$ $=$ 1.
The values are averages of those from each member of the doublet, if unblended 
with other components; otherwise they are from the line that is least affected 
by blending. Horizontal lines in the table
indicate that the component was not detected at that epoch.
The column densities in Table 7 are larger than those derived by assuming 
unsaturated lines by as much as factor of $\sim$3, due to the substantial 
optical depths in many of the components (e.g., see Crenshaw et al. 1998). 

To illustrate the effects of partial covering on the measurements, Table 7 also 
gives column densities for the few instances that we could actually determine 
reliable C$_{los}$ values.
The greatest effect is seen for NGC 3783, where a covering factor of $\sim$0.5 
leads to an increase in the column density by a factor of $\sim$2.
We also note that if the optical depths are high, even a covering factor close 
to one can, if not corrected for, have a significant effect on the column 
densities.
For example, C$_{los}$ = 0.99 and 0.98 for components 1 and 2 in NGC 3516, 
respectively, but the effects on the measured column 
densities are still significant due to the large 
optical depths. In this particular case, we have suggested 
that the measured covering factors are not unity because of grating-scattered 
light, which occurs at the 1 -- 2\% level (Crenshaw et al. 1998). 
Grating-scattered light is almost certainly present at this level in all of the 
GHRS spectra presented here.

We have only a few column densities for the case in which C$_{los}$ has been 
derived. Thus, for the purpose of discussion, we will refer to the column 
densities derived from C$_{los}$ = 1, with the understanding that the true 
column densities may be underestimated by a factor of $\sim$2 or less.

\subsection{Conclusions from the absorption profiles}

In addition to the effects that the covering factors have on column density 
measurements, we can use them to place important constraints on the absorption 
regions. Table 6 shows that C$_{los}^{BLR}$ is always positive, indicating that 
at least a portion of the BLR is occulted 
by each component. This result alone does not provide a tight constraint, 
because the absorber could be closer to the continuum source than the BLR, and 
still intercept a fraction of the emission from the far side of the BLR.
However, we note that there is at least one component in each object with 
C$_{los}^{BLR}$ $\geq$ 0.9, except for NGC 3783. Thus, in four of 
the five Seyfert 1 galaxies in Table 6, there is at least one 
component of the absorbing gas that essentially covers the BLR (or at least the 
high-ionization portion characterized by C~IV and N~V emission). Since the sizes 
of the C~IV emitting regions in these objects are a few 
light days (Wanders et al. 1997, and references therein), the 
absorption regions responsible for these components must be greater than a few 
light days from the continuum source, and greater than a few light days in 
extent (in the plane of the sky).

We define C$_{global}$ to be the fraction of the underlying emission that the 
intrinsic UV absorber (as a whole) covers, averaged over all lines of sight for 
all Seyfert 1 galaxies. Thus,
\begin{equation}
C_{global} = <C_{los}>f,
\end{equation}

where f is the fraction of Seyfert 1 galaxies that show intrinsic absorption, 
which we have determined to be 0.59 for our sample.
For any Seyfert 1 galaxy, C$_{los}$ for the ensemble of absorption components 
must be greater than or equal to the largest value for any single component. We 
take the average of the largest values or lower limits of C$_{los}$ from each 
Seyfert, and find $<$C$_{los}$$>$ $\geq$ 0.86, which leads to  
C$_{global}$ $\geq$ 0.51. We note that these values are estimates, and may be 
affected by the small sample size, as well as biases on the part of the 
observers who selected these Seyfert galaxies.

Thus, as an ensemble, the regions responsible for the UV absorption typically 
cover {\it at least} half of the sky as seen by the central continuum source.
To illustrate this result, we take two extremes: 1) the covering factor could be 
close to one-half in all Seyfert 1 galaxies, or 2) the covering factor could be 
one in the Seyferts with observed absorption and the absorbing gas is not 
present in the other Seyferts. Obviously, the true situation could be somewhere 
in between. In either case, C$_{global}$ provides an important geometric 
constraint for the intrinsic absorber. For example, geometries such
as spherical shells, large pieces of shells or sheets, or cones with large 
opening angles are favored.

Finally, we discuss the column densities in Table 7. We concentrate on the 
values determined for C$_{los}$ $=$ 1 (with the understanding that they 
may be slight underestimates of the true column densities).
There is a substantial range in the column densities of the components, from 0.1 
to 14 x 10$^{14}$ cm$^{-2}$ (the former value is at our detection limit).
By comparison, the column density of C~IV is much higher in a typical BLR cloud: 
N(C~IV) $\geq$ 10$^{18}$ cm$^{-2}$ (Ferland \& Mushotzky 
1982; Ferland et al. 1992).

As we mentioned previously, it is difficult to accomodate the larger values of 
the C~IV column density ($\sim$10$^{15}$ cm$^{-2}$) in X-ray ``warm absorber'' 
models that are characterized by a single ionization parameter.
In addition, we note that in the case of NGC 5548, where multiple components of 
N~V absorption have been measured, the ratio of N(N~V)/N(C~IV) varies 
substantially, ranging from 2.1 to 18.2. This is strong evidence that the 
ionization parameter varies significantly from one kinematic component to the 
next. Combined with the evidence that Si~IV absorption is present in some 
objects and absent in others, and the detection of Mg~II absorption and other 
low ionization lines in only one object (NGC~4151), these results indicate that 
there is a wide range in physical properties, particularly ionization parameter, 
among UV absorbers in different Seyferts and among different components in 
the same Seyfert.  

\section{Variability}

To investigate the variability of the intrinsic UV absorption in our sample, we 
concentrate on the high resolution spectra, since it is difficult to determine 
accurate column densities in the low resolution data. As in the past, when 
absorption variations are detected, thay are found in the column densities of 
the lines and not in the radial velocities. We are limited to the three Seyfert 
1 galaxies with multiple epoch observations at high resolution: NGC 3516, NGC 
3783, and NGC 4151. (Variability studies of other objects with other satellites 
are described in Appendix A.) 

NGC 3783 shows extreme variability in its C~IV absorption lines (Figure 3).
There is no evidence for absorption in the GHRS spectrum on 1993 February 5, but 
a C~IV absorption doublet is present 11 months later (on 1994 January 6) at
$-$560 km s$^{-1}$ (relative to the optical emission). After a subsequent 
interval of 15 months (on 1995 April 15), an additional C~IV doublet is evident 
at a radial velocity of $-$1420 km s$^{-1}$ (there was no significant change in 
the column density of the other component). We reported the first variation in 
Maran et al. (1996), and the second in Crenshaw et al. (1997); in Table 7, we 
give the magnitude of these variations in terms of the derived column densities.
We also note that the single spectrum of the N~V region on 1993 February 21 
(Figure 4) was obtained just 16 days after the first C~IV spectrum, which showed 
no absorption. Since C~IV and N~V are always present together in our 
low-resolution spectra, this {\it suggests} (but does not prove) that the 
absorption varied rapidly over this 16 day interval.

As reported in Crenshaw et al. (1998), we detected no variability in the C~IV 
absorption components of NGC 3516 over a 6-month period, although changes in the 
column density at the $\leq$25\% level cannot be ruled out, due to the fact 
that the lines are saturated. In this case, we know that previous variations of 
the C~IV absorption were due to a broad, blue-shifted component that
disappeared between 1989 October and 1993 February (Koratkar et al. 1996).
For NGC 4151, Weymann et al. (1997) found that the C~IV absorption was constant 
over a 4-year period, with the exception of one transient event. 
We note that over the time span of the HST observations, the lines of 
C~IV and N~V in these 
two Seyferts were saturated, and therefore modest ($\sim$25\%) variations in the 
column densities of these high-ionization lines would not be detected. In NGC 
4151, variations in weaker (and lower ionization) absorption lines have been 
detected with {\it HUT} and {\it ORFEUS}, as
described in Appendix A.

There are two likely sources of absorption variations:
1) changes in the ionization fractions due to continuum variations, and/or 2) 
changes in the total column of gas (e.g., due to bulk motion across the 
line-of-sight).
The former could be identified from a correlation between continuum and 
absorption variations, but obviously we do not have enough observations of NGC 
3783, closely spaced in time, to establish such a correlation.
The best case for a correlation between continuum and absorption variations has 
been established for the neutral hydrogen absorption and low-ionization lines of 
NGC 4151 (Kriss et al. 1996c; Espey et al. 1998).

The best case for bulk motion as a source of variability is the aforementioned 
disappearance of the broad, blueshifted absorption component of C~IV in NGC 
3516. This component has not reappeared in recent observations, despite 
substantial continuum variations, which suggests that the absorbing gas has 
moved out of the line-of-sight. Assuming tangential velocities 
comparable to the largest observed radial velocities, an absorber could move 
across the BLR on a time scale of months to years (Maran et al. 1996). It 
is possible that both continuum variability and bulk motion are responsible for 
absorption variability, and that they could very well be occuring on different 
time scales.

If variations in column density can be attributed to changes in the ionizing 
continuum, then monitoring observations can be used to probe the physical 
conditions in the gas.
The time lag between continuum and absorption variations will provide the 
recombination and/or ionization time scales, and thus the electron density 
(n$_e$) and/or radial location (r) of the gas, respectively (Shields \& Hamann 
1997; Krolik \& Kriss 1997). With only one of these quantities (n$_e$ or r), the 
other can be determined with the assistance of photoionization models, which 
provide the ionization parameter and ionic ratios from the observed column 
densities of different ions. 
For example, Shields \& Hamann (1997) used our first
detection of variability in NGC 3783 to derive lower limits to the electron 
density (n$_e$ $\geq$ 300 cm$^{-3}$) and upper limits to the distance of the 
absorbing gas from the central continuum source (r~$\leq$~10 pc).
Other examples are given in Appendix A, but in every case, only an upper 
limit has been determined for the time lag between continuum and absorption 
variations, which leads to a lower limit on the density and an upper limit on 
the radial distance.

\section{Summary}

We have presented the first systematic study of intrinsic absorption in active 
galaxies that is based on ultraviolet spectra from {\it HST}.
Although the sample is not complete and subject to biases on the part of the 
observers that selected these objects,
we find that intrinsic absorption is much more common than previously believed 
(Ulrich 1988), occuring in more than half (10/17) of the Seyfert 1 galaxies in 
this study. The absorption is due to highly ionized gas that is flowing outward 
from the nucleus.
At high resolution  ($\lambda$/$\Delta\lambda$ $\approx$ 20,000) the absorption 
often breaks up into multiple kinematic 
components, which are likely characterized by different physical conditions, and 
probably different distances from the nucleus.

We have shown that at least some of the absorption components arise from regions 
that are completely outside of the BLR, and that these regions must be at least 
a few light days in size (in the plane of the sky) to occult the BLR.
As an ensemble, the intrinsic absorbers cover a large part of the sky ($\geq$ 
50\%) as seen from the continuum source. We note that these values

The absorption components in our sample remain stable in radial velocity over at 
least several years, but, in one case, show extreme variations in column density 
over one year.
Previous studies have suffered from a lack of spectral or temporal 
resolution, but indicate that the UV absorber lies between the BLR and NLR 
(except for narrow components near the 
systemic velocity of the host galaxy that may arise from that galaxy's halo.)
Intensive monitoring observations (e.g, with {\it HST}) are needed produce 
well-sampled continuum and absorption light curves, in order to explore 
the range of physical characteristics of the absorbing gas, determine its 
origin, and explore its relation to the X-ray warm absorber.

Eight Seyfert 1 galaxies in this study have high-quality {\it HST} and {\it 
ASCA} spectra, and there is a one-to-one correspondence between those objects 
that show intrinsic UV absorption and those that show X-ray warm absorbers. Thus 
these two phenomena are related, but additional studies are needed to determine 
the exact nature of their relationship. Specifically, detailed photoionization 
calculations are needed to match the observed column densities, and thereby test 
the need for multi-zone models. We will address this issue in a companion paper.

\acknowledgments

This research has made use of the NASA/IPAC Extragalactic Database (NED) which 
is operated by the Jet Propulsion Laboratory, California Institute of 
Technology, under contract with the National Aeronautics and Space 
Administration. This research has also made use of NASA's Astrophysics Data 
System Abstract Service. D.M.C. and S.B.K. acknowledge support from NASA grant 
NAG 5-4103. Support for this work was provided by NASA through grant number 
AR-08011.01-96A from the Space Telescope Science Institute, which is operated by 
the Association of Universities for Research in Astronomy, Inc., under NASA 
contract NAS5-26555. 
\clearpage

\appendix

\section{Notes on individual Seyfert 1 galaxies in the sample}

\subsection{WPVS 007 (WPV85 007)}

This relatively unknown narrow-line Seyfert 1 shows variable and ultrasoft X-ray 
emission (Grupe et al. 1995); no information is available on the possibility of 
an X-ray warm absorber.
In addition to deep L$\alpha$, N~V, and C~IV absorption, we have found intrinsic 
Si~IV absorption at the same approximate radial velocity. The FOS observations 
are also described in Goodrich et al. (1998).

\subsection{I Zw 1}

Weak intrinsic absorption was discovered in the FOS spectra of this narrow-line 
Seyfert 1 galaxy by Laor et al. (1997). Although there are several observations, 
there is only one G130H spectrum, which contains the high-ionization absorption 
lines, and therefore no information is available on absorption variability.

\subsection{NGC 3516}

Intrinsic absorption lines of C~IV, N~V, and Si~IV were originally detected in 
{\it IUE} spectra by Ulrich \& Boisson (1983). NGC 3516 also exhibits a strong 
and variable X-ray warm absorber (Kolman et al. 1993; Nandra \& Pounds 1994; 
Kriss et al. 1996b; Mathur, Wilkes, \& Aldcroft 1997; George et al. 1998b). 
Intrinsic O~VI $\lambda\lambda$ 1031.9, 1037.6 absorption has been detected with 
the Hopkins Ultraviolet Telescope ({\it HUT}) by Kriss et al. (1996a). Kriss et 
al. have also detected intrinsic neutral hydrogen absorption (in the Lyman 
series) at a blueshift of $-$400 km s$^{-1}$, relative to the systemic velocity. 

{\it IUE} monitoring found evidence for variability in the C~IV absorption on 
time scales as small as weeks (Voit, Shull, \& Begelman 1987, Walter et al. 
1990; Kolman, Halpern, \& Martin 1993). Walter et al. characterized the C~IV 
absorption as a blend of a narrow stable component in the core of the 
emission-line, and a variable and broad blueshifted component. Kolman et al. 
(1993) found that n$_e$ $\geq$ 10$^{5}$ cm$^{-3}$ and r $\leq$ 9 pc for this 
variable component. Koratkar et al. (1996) show that the variable component 
disappeared between 1989 October and 1993 February, when there were no {\it IUE} 
observations, and has not reappeared. 

Additional information on the FOS observations can be found in Goad et al. 
(1998), and a more detailed description of the GHRS observations is 
presented in Crenshaw et al. (1998). The variable blueshifted component of C~IV 
has not reappeared in the {\it HST} spectra, and there is no evidence for 
variations in the remaining C~IV absorption over the span of {\it HST} 
observations (1995 April - 1996 August).

\subsection{NGC 3783}

Intrinsic L$\alpha$ and C~IV absorption were discovered in the FOS spectrum by 
Reichert et al. (1994), and intrinsic N~V absorption was found by Lu, Savage, \& 
Sembach (1994) in the GHRS spectra of this region. George et al. (1998a) found 
evidence for variability in the X-ray ``warm absorber'' between {\it ASCA} 
observations on 1993 December and 1996 July.

We presented the first two GHRS spectra of the C~IV region in Maran et al. 
(1996), and discovered that the C~IV absorption is extremely variable.
We present the third GHRS spectrum of C~IV in the current paper, and find that
another C~IV component has appeared at a higher blueshift (section 5).
The C~IV absorption detected in the GHRS spectrum on 1994 January 16 at $-$560 
km s$^{-1}$ could be the same component as that seen in the FOS spectrum on 1992 
July 27 (at $-$ 660 km s$^{-1}$), given the weakness of the feature and the 
wavelength uncertainties in the FOS spectra. This component completely 
disappeared on 1993 February 5, but is present in N~V just 16 days later, on 
1993 February 21, suggesting:
1) rapid variability, or 2) unusually high ionization during this time period 
(because in every Seyfert with simultaneous observations, C~IV and N~V 
absorption are always present together). As we discussed earlier, Shields \& 
Hamann (1997) find n$_e$ $\geq$ 300 cm$^{-3}$ and r $\leq$ 10 pc for this 
component.

\subsection{NGC 4151}

A history of the early observations of intrinsic UV absorption is given in the 
Introduction. Previous observations of the complex X-ray absorption are 
described by George et al. (1998b). Observations with {\it HUT} and {\it ORFEUS} 
have extended the UV coverage down to 912 \AA, and have resulted in the 
detection of a number of new absorption lines, including the strong lines of 
C~III $\lambda$977.0, N~III $\lambda$989.8, O~VI $\lambda\lambda$1031.9, 1037.6, 
and the hydrogen Lyman series (Kriss et al. 1992, 1995; Espey et al. 1998). 
Espey et al. show that the C~III$^{*}$ $\lambda$1175 (metastable) absorption 
line varies on a time scale of $\sim$1 day, and thus r $\leq$ 25 pc, similar to 
the value obtained from H~I absorption variations (Krolik \& Kriss 1997). No 
variations were found in the high-ionization absorption lines (C~IV, N~V, O~VI).

The FOS and GHRS spectra were obtained primarily by Weymann et al. (1997).
Due to the large columns, widths, and number of components, the deconvolution of 
the C~IV absorption is much more difficult than for any other object in this 
sample, even at high resolution. Weymann et al. present their deconvolution and 
measurement of individual components, and Kriss (1998) presents a slightly 
different deconvolution; we do not attempt to repeat their analyses here.
Weymann et al. find no evidence for variations in the C~IV absorption in five 
separate observations obtained over $\sim$4 years, except for the appearance of 
a broad absorption feature in the blue wing of C~IV on one occasion. They also 
present evidence for variations in the S~IV absorption in the low resolution 
spectra.

The only other line observed at high resolution is Mg~II (Weymann et al. 1997). 
The Mg~II/C~IV ratio varies greatly from one component to the next, indicating a 
wide range in ionization parameter (Kriss 1998). NGC 4151 is the only 
object that shows Mg~II absorption in our sample, and it would be interesting 
to test the uniqueness of this object, by looking for low-ionization absorption 
in other Seyfert 1 galaxies.

\subsection{NGC 5548}

The intrinsic C~IV absorption was first studied in detail by Shull \& Sachs 
(1993), using data from the 1988 - 1989 {\it IUE} monitoring campaign (Clavel et 
al. 1991). Shull \& Sachs claim that C~IV EW varies on a time scale of 4 days or 
less, and is anticorrelated with the continuum flux at 1570 \AA, indicating that 
n$_e$ $\geq$ 5 x 10$^{5}$ cm $^{-3}$ and r $\leq$ 14 pc. Properties of the X-ray 
warm absorber are given by George et al. (1998b, and references therein).

The FOS observations that we used are from the 1993 {\it HST} monitoring 
campaign (Korista et al. 1995). 
Measurements of the absorption in these data are given by Mathur et al. (1995).
We found no evidence for strong variations in these lines (given the 
difficulties imposed by low resolution), and therefore used the average spectrum 
from this campaign. The GHRS spectra of C~IV (at low and high resolution) were 
obtained by Mathur et al. (1998), and the GHRS spectra of the N~V region are 
also shown in Savage et al. (1997).

\subsection{Mrk 509}

York et al. (1984) found two kinematic components of intrinsic L$\alpha$ 
and C~IV absorption in high-dispersion {\it IUE} spectra. Evidence for an X-ray 
warm absorber is given by Reynolds (1997) and George et al. (1998b).

Our FOS observations found the same two kinematic components as those identified 
by York et al. (1984) $\sim$ 12 years earlier, and we found N~V at these 
velocities as well (Crenshaw et al. 1995). We originally agreed with the 
suggestion of York et al. (1984) that these components arise from extended 
regions of ionized gas in the host galaxy. However the high resolution GHRS 
spectra of L$\alpha$ (also shown in Savage et al. 1997) demonstrates that these 
two components are broad, and are likely to arise close to the nucleus. A 
determination of the location of the absorbers will require variability 
monitoring.

\subsection{II Zw 136}

We find intrinsic L$\alpha$, N~V, and C~IV absorption at two widely separated 
radial velocities in the GHRS low resolution spectra.

\subsection{Akn 564}

The absorption in this narrow-line Seyfert 1 galaxy is similar to that in 
WPVS~007, in that Si~IV absorption is present in addition to strong L$\alpha$, 
N~V, and C~IV. The FOS observations are also presented by Goodrich et al. 
(1998).

\subsection{NGC 7469}

The FOS observations are also shown in Kriss et al. (1998).
We find intrinsic L$\alpha$, N~V, and C~IV absorption at a large blueshift: 
$-$1770 km s$^{-1}$ relative to the emission lines.
George et al. (1998b) found marginal evidence for a warm absorber in {\it ASCA} 
spectra, whereas Reynolds (1997) found no absorption that satisfied his 
criteria.

\section{Additional Seyfert 1 galaxies with possible intrinsic absorption}
Although these Seyferts did not satisfy our selection criteria, they show broad 
L$\alpha$ absorption, indicating possible intrinsic absorption close to the 
nucleus. 

\subsection{III Zw 2}

A noisy FOS G130H spectrum obtained on 1992 January 18 shows moderately broad 
L$\alpha$ absorption in the core of the emission profile, and possible 
intrinsic N~V absorption, near the emission-line redshift of z $=$ 0.0894.

\subsection{Mrk 279}

A high resolution GHRS spectra obtained on 1997 January 16 shows two broad, deep 
L$\alpha$ absorption components, centered at $-$490 km s$^{-1}$ and $+$60 km 
s$^{-1}$ with respect to the emission-line redshift (z $=$ 0.0306).

\subsection{Mrk 290}

A high resolution GHRS spectra obtained on 1997 January 9 shows a broad, deep 
L$\alpha$ absorption component, centered at $-$240 km s$^{-1}$ relative to the 
emission-line redshift (z $=$ 0.0296).

\clearpage

\figcaption[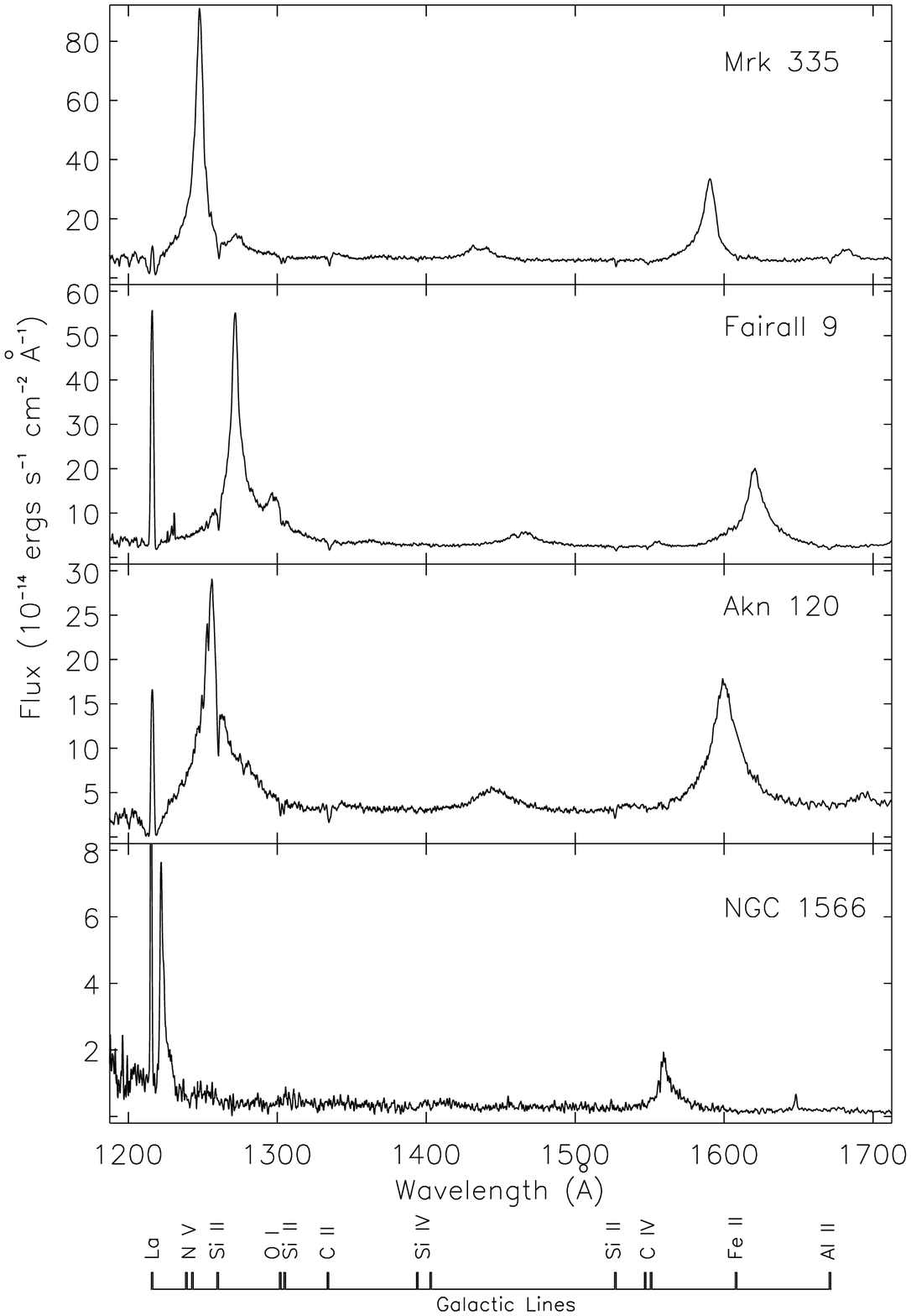, 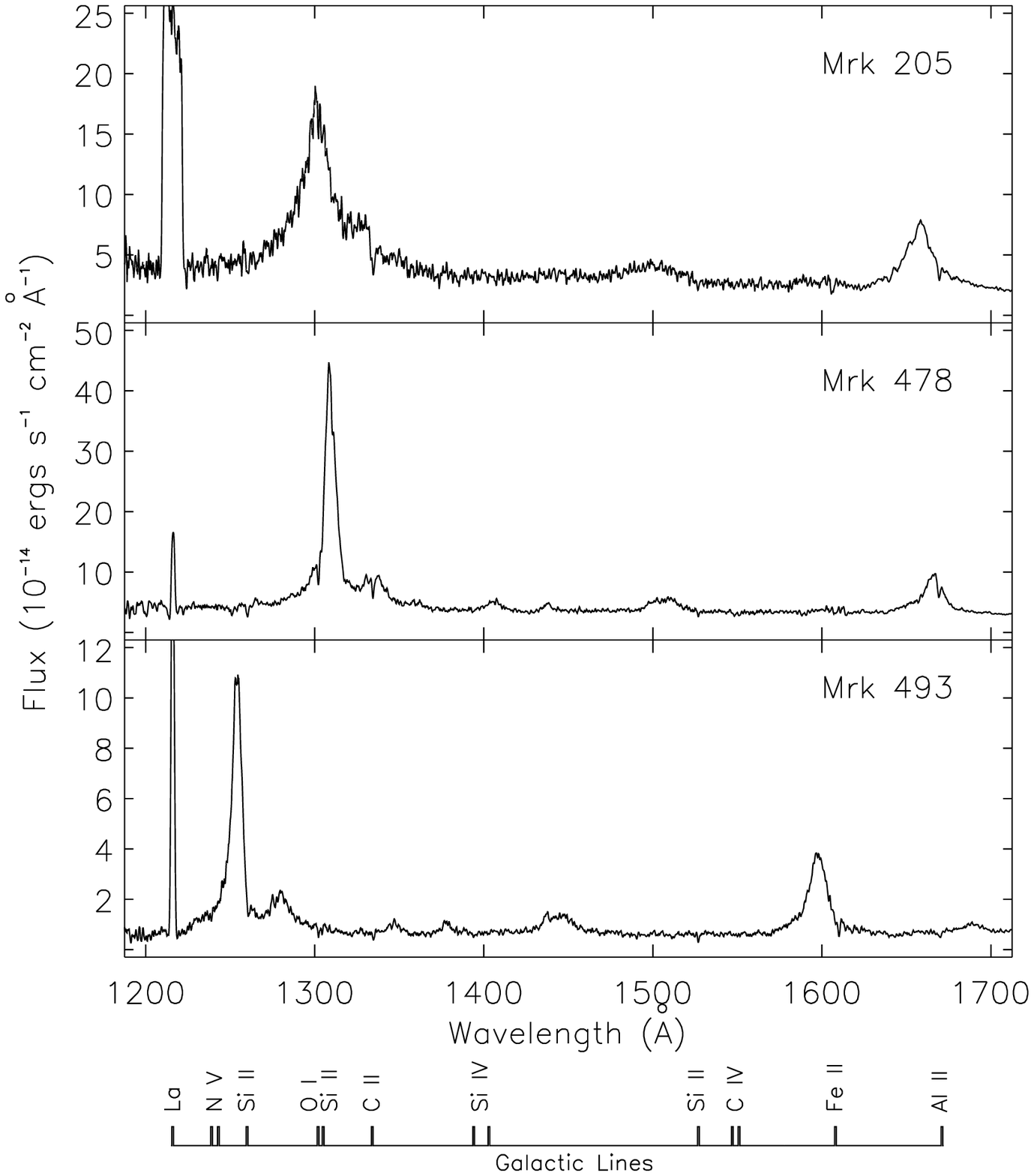]{Far-UV low resolution spectra of seven Seyfert 
1 galaxies that do not show intrinsic UV absorption. The positions of the strong 
Galactic lines are noted at the bottom.}

\figcaption[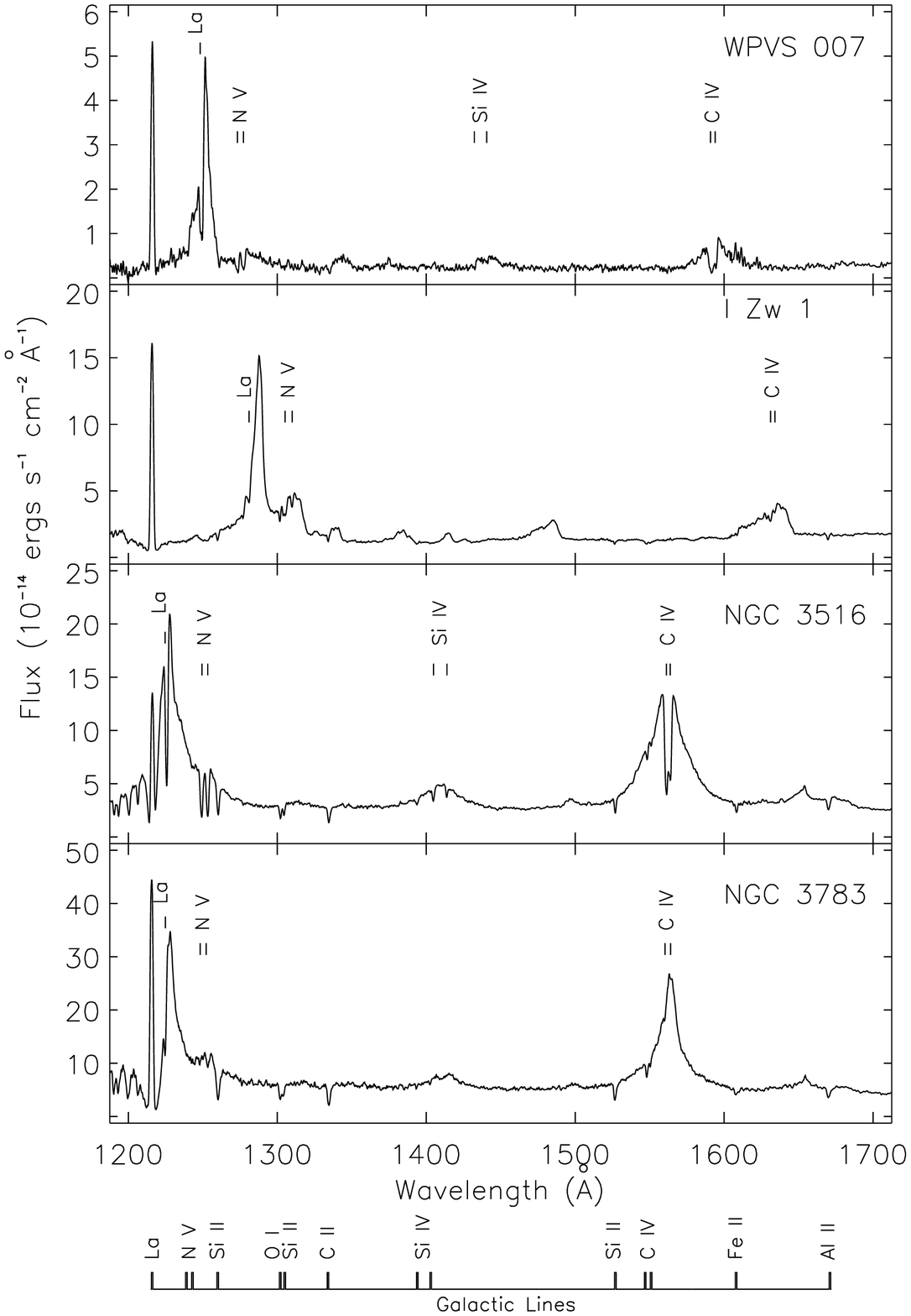, 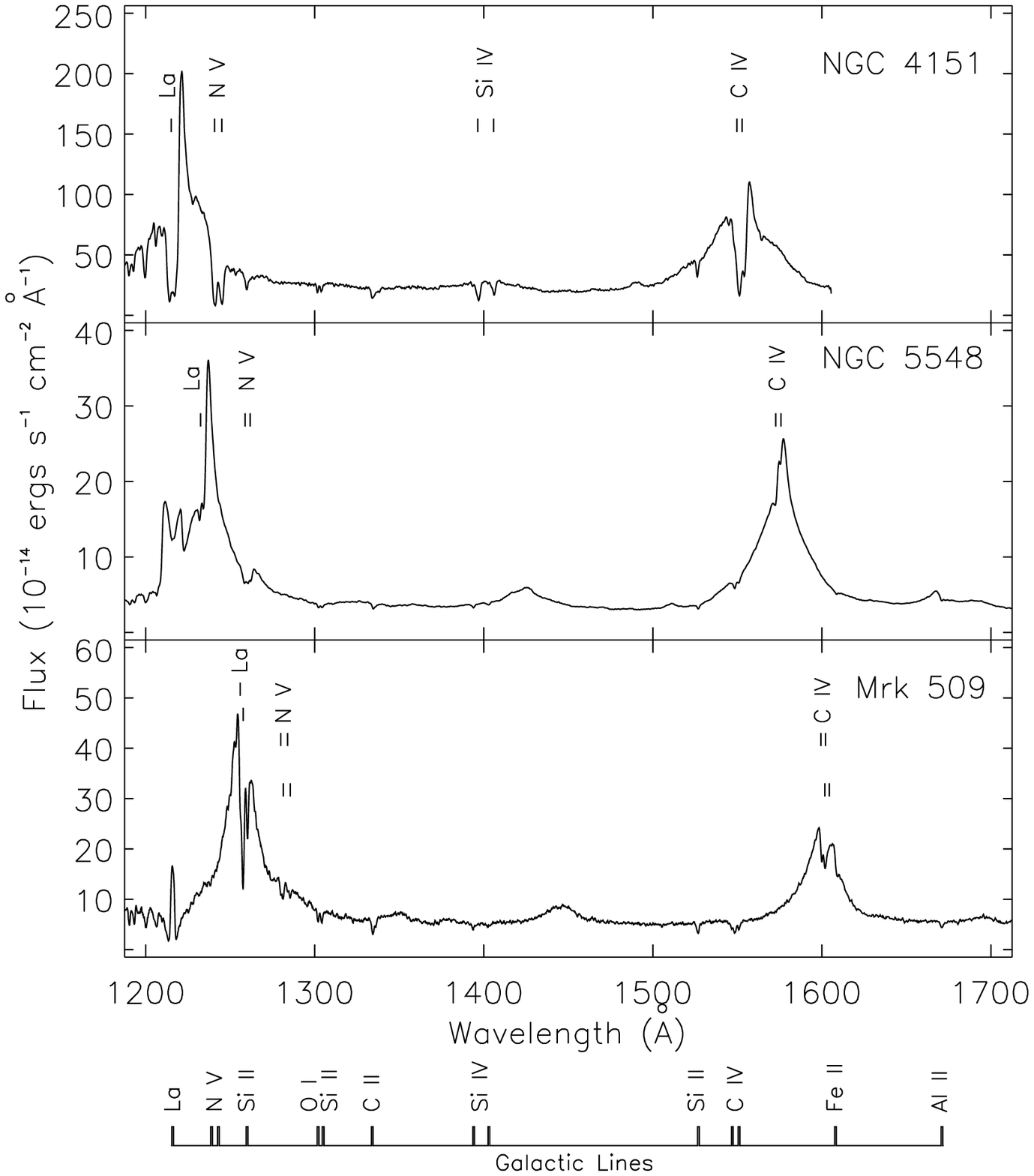, 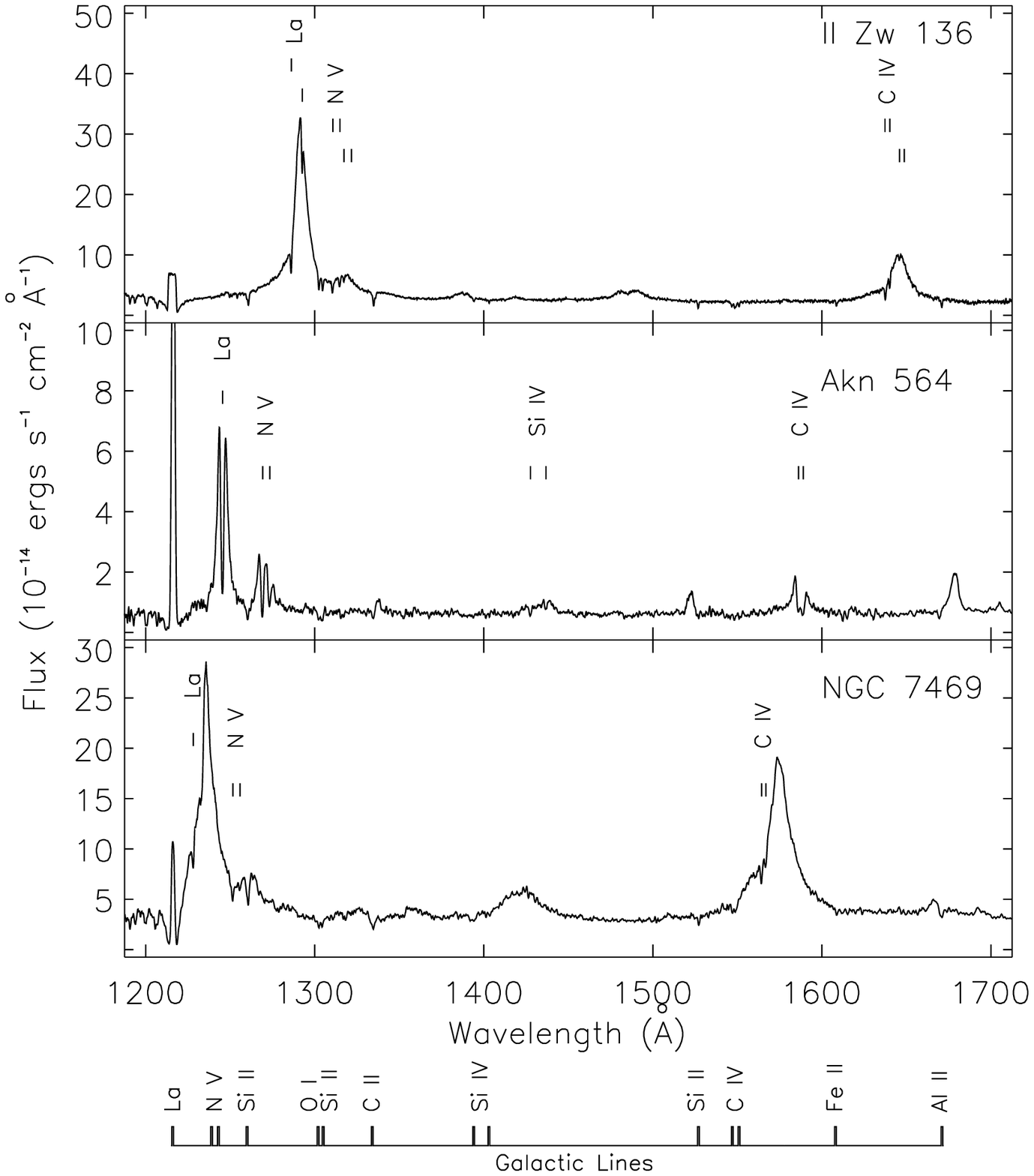]{Far-UV low resolution spectra of 10 
Seyfert 1 galaxies that show intrinsic UV absorption. The positions of the 
intrinsic absorption lines are labeled, and the positions of the strong Galactic 
lines are noted at the bottom.}

\figcaption[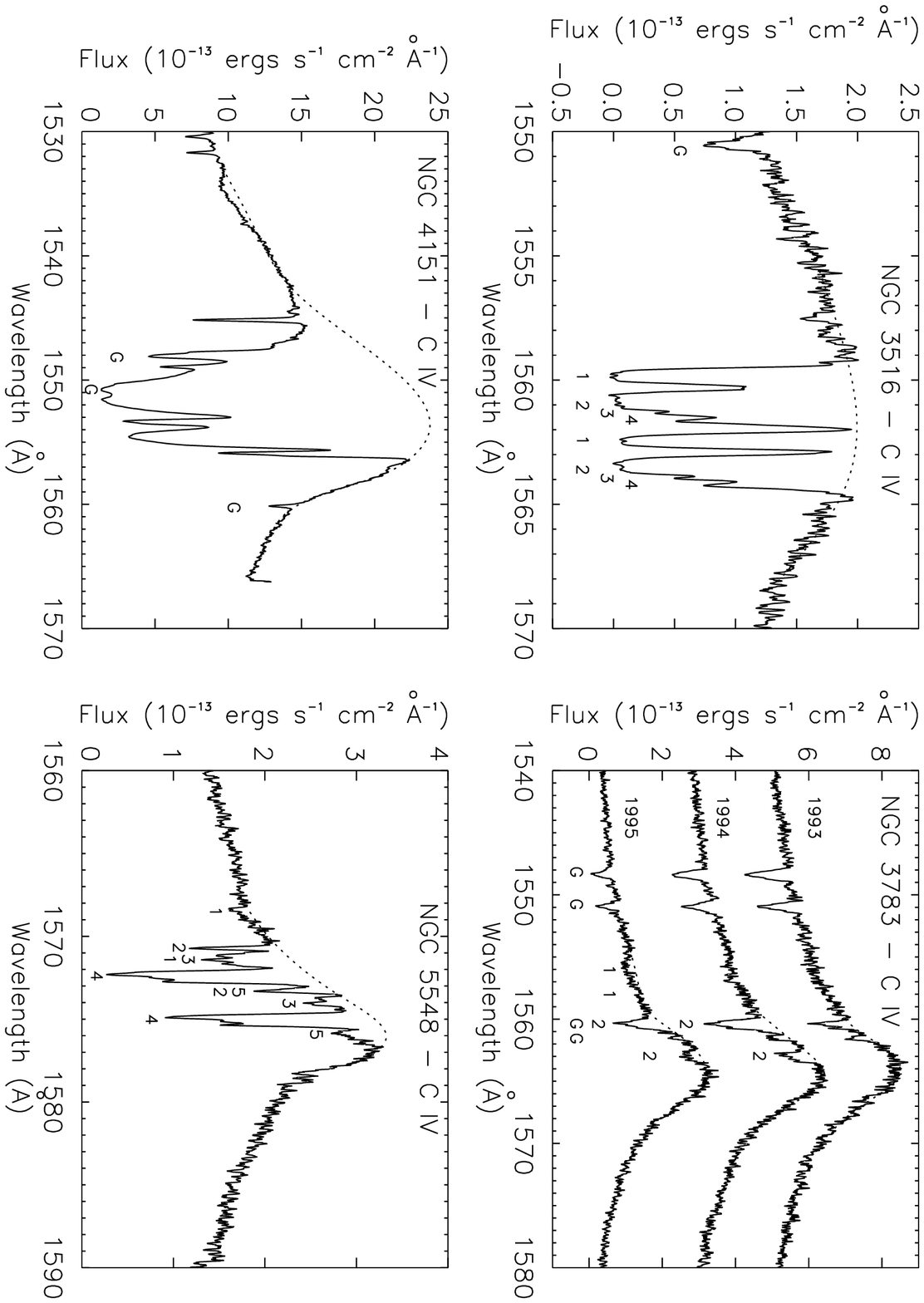]{GHRS high resolution spectra of the C~IV region in four
Seyfert 1 galaxies. The kinematic components are numbered, beginning at the 
shortest wavelengths, and Galactic lines are labled with a ``G''.}

\figcaption[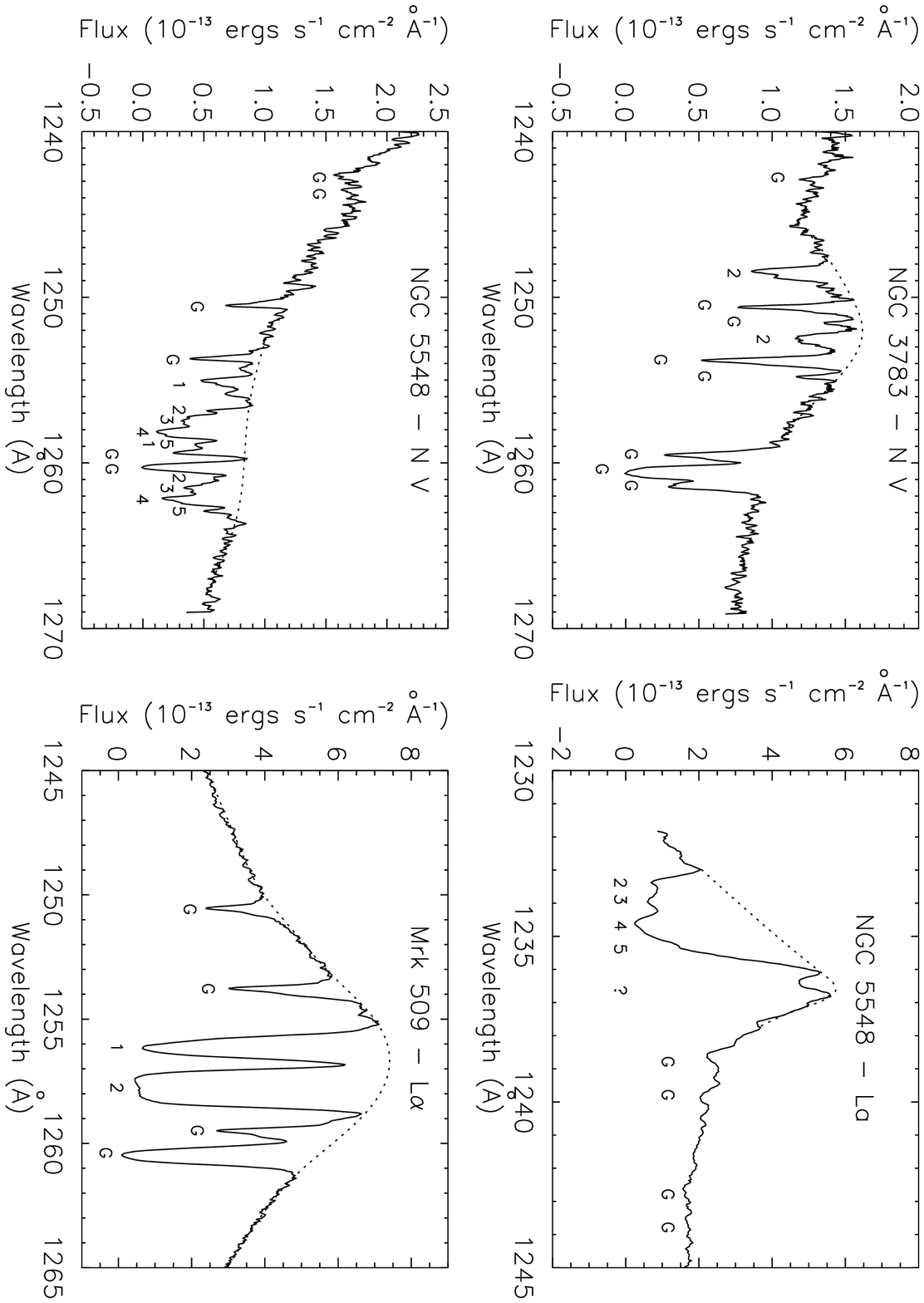]{GHRS high resolution spectra of other lines in this 
sample. The kinematic components are numbered, beginning at the shortest 
wavelengths, and Galactic lines are labled with a ``G''. For the high resolution 
spectrum of Mg~II in NGC 4151, see Weymann et al. (1997).}

\clearpage
\begin{deluxetable}{lllccl}
\tablecolumns{6}
\footnotesize
\tablecaption{Observations \label{tbl-1}}
\tablewidth{0pt}
\tablehead{
\colhead{Name} & \colhead{Inst.} & \colhead{Gratings} &\colhead{Coverage$^{a}$} 
&\colhead{Aper.$^{b}$ ($''$)} &\colhead{Observation Date (UT)}
}
\startdata
\multicolumn{6}{c}{Low resolution spectra}\nl
\tableline
Mrk 335    &FOS &G130H, G190H, G270H  &1150 - 3280  &1.0 &1994 December 16 \\
WPVS 007   &FOS &G130H, G190H, G270H  &1150 - 3280  &1.0 &1996 July 30 \\
I Zw 1 (1) &FOS &G270H                &2220 - 3280  &1.0 &1992 July 29 \\
I Zw 1 (2) &FOS &G190, G270H       &1575 - 3280  &4.3 &1993 January 16 \\
I Zw 1 (3) &FOS &G130H            &1150 - 1605  &1.0 &1994 February 13 \\
I Zw 1 (4) &FOS &G190H, G270H     &1575 - 3280  &1.0 &1994 September 14 \\
Fairall 9  &FOS &G130H, G190H, G270H  &1150 - 3280  &1.0 &1993 January 22 \\
NGC 1566   &FOS &G130H, G190H, G270H  &1150 - 3280  &1.0 &1994 February 8, 11 
\\  
Akn 120    &FOS &G130H, G190H, G270H  &1150 - 3280  &1.0 &1995 1995 July 29 \\
NGC 3516 (1) &FOS &G130H, G190H, G270H  &1150 - 3280  &1.0 &1995 December 30 \\
NGC 3516 (2) &FOS &G130H, G190H, G270H  &1150 - 3280  &1.0 &1996 February 21 \\
NGC 3516 (3) &FOS &G130H, G190H, G270H  &1150 - 3280  &1.0 &1996 April 13 \\
NGC 3516 (4) &FOS &G130H, G190H  G270H  &1150 - 3280  &1.0 &1996 November 28 \\
NGC 3516 (5) &FOS &G130H, G190H, G270H  &1150 - 3280  &1.0 &1996 August 14 \\
NGC 3783     &FOS &G130H, G190H         &1150 - 2330  &1.0 &1992 July 27 \\    
NGC 4151 (1) &FOS &G130H                &1150 - 1605  &1.0 &1992 June 22 - 27 
\\  
NGC 4151 (2) &FOS &G130H                &1150 - 1605  &0.25x2.0 &1992 July 3 \\  
NGC 4151 (3) &FOS &G270H                &2220 - 3280  &0.3 &1993 April 22 \\  
NGC 4151 (4) &FOS &G130H                &1150 - 1605  &0.3 &1996 March 11 \\  
Mrk 205 (1)  &FOS &G190H, G270H         &1575 - 3280  &0.25x2.0 &1991 July 11 
\\ 
Mrk 205 (2)  &FOS &G130H                &1150 - 1605  &4.3      &1993 July 18 
\\ 
NGC 5548 (1) &FOS &G130H, G190H, G270H  &1150 - 3280  &1.0 &1992 July 5 \\ 
NGC 5548 (2) &FOS &G130H, G190H    &1150 - 2330  &4.3 &1993 April 19 - May 27 
\\ 
NGC 5548 (3) &GHRS &G140L             &1415 - 1715  &2.0 &1996 August 24 \\
Mrk 478    &FOS &G130H, G190H, G270H  &1150 - 3280  &1.0 &1996 December 5 \\
Mrk 493    &FOS &G130H, G190H, G270H  &1150 - 3300  &1.0 &1996 September 4 \\
Mrk 509    &FOS &G130H, G190H, G270H  &1150 - 3300  &1.0 &1992 June 21 \\ 
\tablebreak
II Zw 136  &GHRS &G140L               &1153 - 1740  &2.0 &1995 July 24 \\ 
Akn 564    &FOS &G130H, G190H, G270H  &1150 - 3280  &1.0 &1996 May 23 \\
NGC 7469   &FOS &G130H, G190H, G270H  &1150 - 3300  &1.0 &1996 June 18 \\ 
\tableline
\multicolumn{6}{c}{}\nl
\multicolumn{6}{c}{High resolution spectra}\nl
\tableline
NGC 3516 (1)  &GHRS &G160M         &C~IV  &2.0         &1995 April 25 \\
NGC 3516 (2)  &GHRS &G160M         &C~IV  &2.0         &1995 October22 \\
NGC 3783 (1)  &GHRS &G160M         &N~V   &2.0         &1993 February 21 \\
NGC 3783 (2)  &GHRS &G160M         &C~IV  &2.0         &1993 February 5 \\
NGC 3783 (3)  &GHRS &G160M         &C~IV  &2.0         &1994 January 16 \\
NGC 3783 (4)  &GHRS &G160M         &C~IV  &2.0         &1995 April 11 \\
NGC 4151 (1)  &GHRS &G160M         &C~IV  &2.0         &1992 July 4 \\
NGC 4151 (2)  &GHRS &G160M, G270M  &C~IV, Mg~II  &2.0  &1994 January 3 \\
NGC 4151 (3)  &GHRS &G160M, G270M  &C~IV, Mg~II  &2.0  &1994 October 28 - 29 \\
NGC 4151 (4)  &GHRS &G160M, G270M  &C~IV, Mg~II  &2.0  &1996 March 11 \\
NGC 5548 (1)  &GHRS &G160M         &L$\alpha$, N~V  &2.0  &1996 February 17 \\
NGC 5548 (2)  &GHRS &G160M         &C~IV         &2.0  &1996 August 24 \\
Mrk 509       &GHRS &G160M         & L$\alpha$         &2.0  &1993 May 22 \\      
\tablenotetext{a}{For the low resolution spectra, coverage is given in \AA.
For the high resolution spectra, coverage is specified by the emission-line 
profile that is present.}
\tablenotetext{b}{The shapes and exact dimensions of the apertures, before and 
after the installation of COSTAR in 1993 December, are given in Keyes et al. 
1995 and Soderblom et al. 1995.}
\enddata
\end{deluxetable}

\clearpage
\begin{deluxetable}{lcccc}
\tablecolumns{5}
\footnotesize
\tablecaption{Redshifts and intrinsic absorption detections \label{tbl-3}}
\tablewidth{0pt}
\tablehead{
\colhead{Name} & \colhead{z (H I)} & \colhead{z (optical)$^{a}$} &
\colhead{Intrinsic UV} & \colhead{X-ray Warm} \\
\colhead{} & \colhead{} & \colhead{} &
\colhead{Absorption?} & \colhead{Absorber?} \\
}
\startdata
Mrk 335          &         &0.02564 &No  &No \\
WPVS 007$^{b}$   &         &0.02882 &Yes &   \\
I Zw 1$^{b}$     &0.06108  &0.06072 &Yes &   \\
Fairall 9        &         &0.04614 &No  &No \\
NGC 1566         &0.00499  &0.00483 &No  &   \\
Akn 120          &         &0.03312 &No  &   \\
NGC 3516         &         &0.00875 &Yes &Yes \\
NGC 3783         &         &0.00976 &Yes &Yes \\
NGC 4151         &0.00332  &0.00319 &Yes &Yes \\
Mrk 205          &         &0.07085 &No  &   \\
NGC 5548         &0.01717  &0.01676 &Yes &Yes \\
Mrk 478          &         &0.07905 &No  &   \\
Mrk 493$^{b}$    &0.03148  &0.03131 &No  &   \\
Mrk 509          &         &0.03440 &Yes &Yes \\
II Zw 136        &0.06298  &0.06305$^{c}$ &Yes &   \\
Akn 564$^{b}$    &0.02467  &0.02499 &Yes &  \\
NGC 7469         &0.01640  &0.01616 &Yes &Yes?$^{d}$ \\
\tablenotetext{a}{Primarily from the narrow [O III] lines.}
\tablenotetext{b}{Narrow-line Seyfert 1.}
\tablenotetext{c}{from Vrtilek \& Carleton (1985).}
\tablenotetext{d}{Yes according to George et al. (1998); no according to
Reynolds (1997).}
\enddata
\end{deluxetable}

\begin{deluxetable}{lcclc}
\tablecolumns{5}
\footnotesize
\tablecaption{Direct measurements of low resolution spectra \label{tbl-4}}
\tablewidth{0pt}
\tablehead{
\colhead{Name} & \colhead{$\lambda_{obs}$} & \colhead{EW} &
\colhead{Line} & \colhead{v$_{r}$} \\
\colhead{} & \colhead{(\AA)} &
\colhead{(\AA)} & \colhead{(Component)} & \colhead{(km s$^{-1}$)}\\
}
\startdata
WPVS 007  &1249.13 &1.69$\pm$0.19 &L$\alpha$		  &$-$390 \\
          &1273.08 &1.80$\pm$0.34 &N V $\lambda$1238.8    &$-$348 \\
          &1277.22 &1.76$\pm$0.29 &N V $\lambda$1242.8    &$-$336 \\
          &1432.20 &1.06$\pm$0.32 &Si IV $\lambda$1393.8  &$-$369 \\
          &1440.48 &0.72$\pm$0.33 &Si IV $\lambda$1402.8  &$-$366 \\
          &1590.76 &2.59$\pm$0.41 &C IV $\lambda$1548.2   &$-$399 \\
          &1593.85 &2.29$\pm$0.23 &C IV $\lambda$1550.8   &$-$311 \\
 & & & &\\
I Zw 1    &1280.88 &0.49$\pm$0.07 &L$\alpha$		  &$-$2122 \\
          &1305.08 &0.68$\pm$0.07 &N V $\lambda$1238.8    &$-$2167 \\
          &1309.90 &0.27$\pm$0.05 &N V $\lambda$1242.8    &$-$2017 \\
          &1631.07 &0.63$\pm$0.11 &C IV $\lambda$1548.2   &$-$2155 \\
          &1633.95 &0.26$\pm$0.10 &C IV $\lambda$1550.8   &$-$2123 \\
 & & & &\\
NGC 3516$^a$  &1225.60 &1.87$\pm$0.07 &L$\alpha$		  &$-$174 \\
          &1249.29 &1.62$\pm$0.03 &N V $\lambda$1238.8    &$-$90 \\
          &1253.20 &1.54$\pm$0.06 &N V $\lambda$1242.8    &$-$114 \\
          &1404.82 &0.47$\pm$0.06 &Si IV $\lambda$1393.8  &$-$243 \\
          &1413.82 &0.31$\pm$0.06 &Si IV $\lambda$1402.8  &$-$261 \\
          &1561.08 &1.73$\pm$0.11 &C IV $\lambda$1548.2   &$-$128 \\
          &1563.69 &1.55$\pm$0.10 &C IV $\lambda$1550.8   &$-$126 \\
 & & & &\\
NGC 3783  &1224.79 &0.81$\pm$0.07 &L$\alpha$		  &$-$677 \\
          &1248.07 &0.18$\pm$0.03 &N V $\lambda$1238.8    &$-$686 \\
          &1253.1  &-----	  &N V $\lambda$1242.8,   &----- \\
          &	   &		  &Gal. S II $\lambda$1253.8 & \\
          &1560.17 &0.12$\pm$0.04 &C IV $\lambda$1548.2   &$-$608 \\
          &1562.93 &0.07$\pm$0.03 &C IV $\lambda$1550.8   &$-$575 \\
\tablebreak
NGC 4151$^a$  &1214.76 &-----         &L$\alpha$, Gal. L$\alpha$  &----- \\
          &1240.63 &3.04$\pm$0.32 &N V $\lambda$1238.8    &$-$519 \\
          &1244.86 &2.49$\pm$0.22 &N V $\lambda$1242.8    &$-$459 \\
          &1396.32 &1.92$\pm$0.32 &Si IV $\lambda$1393.8  &$-$405 \\
          &1405.73 &1.19$\pm$0.19 &Si IV $\lambda$1402.8  &$-$323 \\
          &1549.58 &3.76$\pm$0.51 &C IV $\lambda$1548.2   &$-$689 \\
          &1553.73 &2.84$\pm$0.39 &C IV $\lambda$1550.8   &$-$384 \\
 & & & &\\
NGC 5548$^a$  &1234.45 &0.73$\pm$0.06 &L$\alpha$		  &$-$393 \\
          &1258.71 &-----	  &N V $\lambda$1238.8,   &----- \\
          &	   &		  &Gal. Si II $\lambda$1260.4	& \\
          &1261.66 &-----	  &N V $\lambda$1242.8,   &----- \\
          &	   &		  &Gal. Si II $\lambda$1260.4	& \\
          &1572.60 &C IV $\lambda$1548.2   &0.23$\pm$0.04  &$-$300 \\
          &1575.74 &C IV $\lambda$1550.8   &0.12$\pm$0.02  &$-$200 \\
 & & & &\\
Mrk 509   &1255.80 &0.48$\pm$0.11 &L$\alpha$ (1)	     &$-$417 \\
          &1257.64 &1.54$\pm$0.04 &L$\alpha$ (2)	     &$+$36 \\
          &1279.77 &0.25$\pm$0.05 &N V $\lambda$1238.8 (1)   &$-$399 \\
          &1281.27 &0.50$\pm$0.06 &N V $\lambda$1238.8 (2)   &$-$36 \\
          &1283.81 &0.17$\pm$0.06 &N V $\lambda$1242.8 (1)   &$-$420 \\
          &1285.39 &0.28$\pm$0.05 &N V $\lambda$1242.8 (2)   &$-$40 \\
          &-----$^{b}$ &-----     &C IV $\lambda$1548.2 (1),  &----- \\
          &   &                   &C IV $\lambda$1550.8 (1),  & \\
          &   &                   &C IV $\lambda$1548.2 (2),  & \\
          &   &                   &C IV $\lambda$1550.8 (2) & \\
\tablebreak
II Zw 136 &1286.07 &0.38$\pm$0.04 &L$\alpha$ (1)	     &$-$1541 \\
          &1292.53 &0.17$\pm$0.02 &L$\alpha$ (2)	     &$+$52 \\
          &1310.57 &0.37$\pm$0.07 &N V $\lambda$1238.8 (1)   &$-$1539 \\
          &1314.75 &0.26$\pm$0.05 &N V $\lambda$1242.8 (1)   &$-$1546 \\
          &1317.19 &0.14$\pm$0.03 &N V $\lambda$1238.8 (2)   &$+$63 \\
          &1321.58 &0.09$\pm$0.03 &N V $\lambda$1242.8 (2)   &$+$101 \\
          &1637.62 &0.46$\pm$0.06 &C IV $\lambda$1548.2 (1)  &$-$1587 \\
          &1640.12 &0.29$\pm$0.05 &C IV $\lambda$1550.8  (1) &$-$1629 \\
          &1645.81 &0.10$\pm$0.03 &C IV $\lambda$1548.2 (2)  &$-$1 \\
          &1648.61 &0.06$\pm$0.03 &C IV $\lambda$1550.8  (2) &$+$12 \\
 & & & &\\
Akn 564   &1245.40 &1.85$\pm$0.13 &L$\alpha$		  &$-$159 \\
          &1269.13 &1.89$\pm$0.24 &N V $\lambda$1238.8    &$-$157 \\
          &1273.26 &1.49$\pm$0.26 &N V $\lambda$1242.8    &$-$144 \\
          &1427.54 &0.56$\pm$0.18 &Si IV $\lambda$1393.8  &$-$224 \\
          &1436.71 &0.44$\pm$0.15 &Si IV $\lambda$1402.8  &$-$240 \\
          &1586.14 &1.69$\pm$0.14 &C IV $\lambda$1548.2   &$-$144 \\
          &1588.89 &1.63$\pm$0.17 &C IV $\lambda$1550.8   &$-$123 \\
 & & & &\\
NGC 7469  &1227.97 &0.38$\pm$0.07 &L$\alpha$		  &$-$1810 \\
          &1251.53 &0.33$\pm$0.07 &N V $\lambda$1238.8    &$-$1814 \\
          &1255.70 &0.16$\pm$0.08 &N V $\lambda$1242.8    &$-$1732 \\
          &1564.24 &0.26$\pm$0.07 &C IV $\lambda$1548.2   &$-$1738 \\
          &1566.80 &0.16$\pm$0.06 &C IV $\lambda$1550.8   &$-$1745 \\
\tablenotetext{a}{No evidence for variability; average values are given.}
\tablenotetext{b}{A deconvolution
of the components in given in Crenshaw et al. (1995).}
\enddata
\end{deluxetable}

\begin{deluxetable}{lcclcc}
\tablecolumns{6}
\footnotesize
\tablecaption{Direct measurements of high resolution spectra$^{a}$
\label{tbl-5}}
\tablewidth{0pt}
\tablehead{
\colhead{Name} & \colhead{$\lambda_{obs}$} & \colhead{EW} &
\colhead{Line} & \colhead{v$_{r}$} & \colhead{FWHM} \\
\colhead{(Epoch)} & \colhead{(\AA)} & \colhead{(\AA)} &
\colhead{(Component)} & \colhead{(km s$^{-1}$)}& \colhead{(km s$^{-1}$)}
}
\startdata
NGC 3516 (1) &1559.87 &0.75$\pm$0.04 &C IV $\lambda$1548.2 (1) &$-$375  &138 \\
             &1561.04 &1.11$\pm$0.04 &C IV $\lambda$1548.2 (2) &$-$148  &220 \\
             &1561.34 &0.04$\pm$0.02 &C IV $\lambda$1548.2 (3) &$-$90   &~17 \\
             &1561.63 &0.08$\pm$0.02 &C IV $\lambda$1548.2 (4) &$-$34   &~29 \\
             &1562.44 &0.59$\pm$0.02 &C IV $\lambda$1550.8 (1) &$-$379  &121 \\
             &1563.63 &0.99$\pm$0.02 &C IV $\lambda$1550.8 (2) &$-$149  &207 \\
             &1563.96 &0.04$\pm$0.02 &C IV $\lambda$1550.8 (3) &$-$85   &~17 \\
             &1564.23 &0.07$\pm$0.02 &C IV $\lambda$1550.8 (4) &$-$33   &~25 \\
NGC 3516 (2) &1559.87 &0.68$\pm$0.03 &C IV $\lambda$1548.2 (1) &$-$375  &127 \\
             &1561.06 &1.16$\pm$0.05 &C IV $\lambda$1548.2 (2) &$-$144  &207 \\
             &1561.35 &0.06$\pm$0.04 &C IV $\lambda$1548.2 (3) &$-$88   &~23 \\
             &1561.65 &0.09$\pm$0.03 &C IV $\lambda$1548.2 (4) &$-$30   &~34 \\
             &1562.44 &0.56$\pm$0.03 &C IV $\lambda$1550.8 (1) &$-$379  &119 \\
             &1563.62 &1.03$\pm$0.04 &C IV $\lambda$1550.8 (2) &$-$151  &190 \\
             &1563.95 &0.03$\pm$0.02 &C IV $\lambda$1550.8 (3) &$-$87   &~23 \\
             &1564.26 &0.06$\pm$0.03 &C IV $\lambda$1550.8 (4) &$-$27   &~36 \\
NGC 3783 (1) &1248.58 &0.28$\pm$0.04 &N V $\lambda$1238.8 (2)  &$-$563  &180 \\
	     &1252.57 &0.19$\pm$0.03 &N V $\lambda$1242.8 (2)  &$-$569  &179 \\
NGC 3783 (3)$^{b}$ &1560.46 &----- &C IV $\lambda$1548.2 (2), &-----   &-----\\
             &        &              &Gal. C I $\lambda$1560.3 &        &    \\
             &1563.01 &0.14$\pm$0.01 &C IV $\lambda$1550.8 (2) &$-$561  &134 \\
NGC 3783 (4) &1556.07 &0.39$\pm$0.08 &C IV $\lambda$1548.2 (1) &$-$1403 &370 \\
             &1558.44 &0.14$\pm$0.08 &C IV $\lambda$1550.8 (1) &$-$1442 &320 \\
             &1560.44 &-----        &C IV $\lambda$1548.2 (2), &-----  &-----\\
             &        &              &Gal. C I $\lambda$1560.3 &        &    \\
             &1563.01 &0.13$\pm$0.03 &C IV $\lambda$1550.8 (2) &$-$560  &152 \\
\tablebreak 
NGC 5548 (1) &1255.32 &0.34$\pm$0.05 &N V $\lambda$1238.8 (1)  &$-$1031 &114 \\
             &1256.84 &0.09$\pm$0.03 &N V $\lambda$1238.8 (2)  &$-$1663 & 52 \\
             &1257.40 &0.41$\pm$0.06 &N V $\lambda$1238.8 (3)  &$-$528 &-----\\
             &1258.18 &0.59$\pm$0.04 &N V $\lambda$1238.8 (4)  &$-$339  &188 \\
             &1258.90 &0.22$\pm$0.04 &N V $\lambda$1238.8 (5), &-----  &-----\\
             &        & 	     &N V $\lambda$1242.8 (1)  & & \\
             &1260.95 &0.09$\pm$0.03 &N V $\lambda$1242.8 (2)  &$-$647 &-----\\
             &1261.54 &0.36$\pm$0.03 &N V $\lambda$1242.8 (3)  &$-$504 &-----\\
             &1262.25 &0.47$\pm$0.04 &N V $\lambda$1242.8 (4)  &$-$373 &-----\\
             &1262.91 &0.09$\pm$0.02 &N V $\lambda$1242.8 (5)  &$-$174  &57  \\
NGC 5548 (2) &1568.53 &0.05$\pm$0.02 &C IV $\lambda$1548.2 (1) &$-$1088 &-----\\
             &1570.72 &0.09$\pm$0.01 &C IV $\lambda$1548.2 (2) &$-$662  &38  \\
             &1571.41 &0.29$\pm$0.03 &C IV $\lambda$1548.2 (3), &-----  &-----\\
             &        &              &C IV $\lambda$1550.8 (1)  &       &    \\
             &1572.43 &0.62$\pm$0.04 &C IV $\lambda$1548.2 (4) &$-$333  &145 \\
             &1573.27 &0.09$\pm$0.02 &C IV $\lambda$1548.2 (5)  &-----  &-----\\
             &        &              &C IV $\lambda$1550.8 (2)  &       &    \\
             &1574.07 &0.10$\pm$0.03 &C IV $\lambda$1550.8 (3) &$-$521  &90  \\
             &1575.06 &0.45$\pm$0.04 &C IV $\lambda$1550.8 (4) &$-$330  &135 \\
             &1575.95 &0.07$\pm$0.03 &C IV $\lambda$1550.8 (5) &$-$156  &90  \\
Mrk 509      &1256.15 &0.83$\pm$0.03 &L$\alpha$ (1)            &$-$330  &215 \\
             &1257.78 &1.52$\pm$0.02 &L$\alpha$ (2)            &$+$72   &367 \\
\tablenotetext{a}{Measurements for NGC 4151 are given by Weymann et al. 1997.}
\tablenotetext{b}{No intrinsic absorption was detected in epoch 2 (Maran et al. 
1996).}
\enddata
\end{deluxetable}

\begin{deluxetable}{lcc}
\tablecolumns{3}
\footnotesize
\tablecaption{Kinematic components in the low resolution spectra \label{tbl-6}}
\tablewidth{0pt}
\tablehead{
\colhead{Name} & \colhead{Component} & \colhead{v$_{r}$}  \\
\colhead{} & \colhead{} & \colhead{(km s$^{-1}$)} \\
}
\startdata
WPVS 007    &1& $-$360$\pm$31  \\
I Zw 1      &1&$-$2117$\pm$59  \\
NGC 3516    &1& $-$162$\pm$66  \\
NGC 3783    &1& $-$657$\pm$43  \\
NGC 4151    &1& $-$463$\pm$129 \\
NGC 5548    &1& $-$298$\pm$96  \\
Mrk 509     &1& $-$412$\pm$11  \\
            &2&  $-$13$\pm$42  \\
II Zw 136   &1& $-$1568$\pm$39 \\
            &2& $+$45$\pm$41   \\
Akn 564     &1& $-$170$\pm$44  \\
NGC 7469    &1&$-$1768$\pm$40  \\
\enddata
\end{deluxetable}

\begin{deluxetable}{lccclll}
\tablecolumns{7}
\footnotesize
\tablecaption{Kinematic components and covering factors in the high resolution 
spectra \label{tbl-7}}
\tablewidth{0pt}
\tablehead{
\colhead{Name} & \colhead{Component} & \colhead{v$_{r}$} & \colhead{FHWM} &
\colhead{~C$_{los}$} & \colhead{C$_{los}^{BLR}$} & \colhead{Line$^{a}$}\\
\colhead{} & \colhead{} & \colhead{(km s$^{-1}$)} &
\colhead{(km s$^{-1}$)} & \colhead{} & \colhead{} & \colhead{}\\
}
\startdata
NGC 3516    &1 & $-$377$\pm$2  &126$\pm$9  &~0.99$\pm$0.01 &0.99      &(C~IV)\\
            &2 & $-$148$\pm$3  &206$\pm$12 &~0.98$\pm$0.01 &0.97       &(C~IV)\\
	    &3 &  $-$88$\pm$2  & 20$\pm$4  &$\geq$0.83    &$\geq$0.79 &(C~IV)\\
	    &4 &  $-$31$\pm$3  & 31$\pm$5  &$\geq$0.74    &$\geq$0.67 &(C~IV)\\
NGC 3783    &1 &$-$1422$\pm$28 &345$\pm$35 &$\geq$0.21  &$\geq$$-$0.45 &(C~IV)\\
            &2 & $-$563$\pm$4  &161$\pm$22 &~~0.52$\pm$0.25 &0.14       &(N~V)\\
NGC 4151$^{b}$ &--- &---        &---        &$\geq$0.94    &$\geq$0.93 &(C~IV)\\
NGC 5548    &1 &$-$1060$\pm$40 &114$\pm$10 &$\geq$0.52    &$\geq$0.23 &(N~V)\\
            &2 & $-$655$\pm$11 & 45$\pm$10 &$\geq$0.76    &$\geq$0.72 
&(L$\alpha$)\\
            &3 & $-$518$\pm$12 & 90$\pm$10 &$\geq$0.81  &$\geq$0.79 
&(L$\alpha$)\\
	    &4 & $-$344$\pm$20 &156$\pm$28 &~0.96$\pm$0.04 &0.95   &(C~IV)\\     
	    &5 & $-$165$\pm$13 & 74$\pm$23 &$\geq$0.69  &$\geq$0.66 
&(L$\alpha$)\\
Mrk 509	    &1 & $-$330$\pm$10 &215$\pm$10 &$\geq$0.91  &$\geq$0.90 
&(L$\alpha$)\\
	    &2 &  $+$72$\pm$10 &367$\pm$10 &$\geq$0.93  &$\geq$0.92 
&(L$\alpha$)\\
\tablenotetext{a}{Line from which the covering factors were determined.}
\tablenotetext{b}{From component ``E'' in Weymann et al. (1997) -- see this 
reference for additional measurements.}
\enddata
\end{deluxetable}

\begin{deluxetable}{lrcrr}
\tablecolumns{5}
\footnotesize
\tablecaption{Column densities for high resolution spectra \label{tbl-8}}
\tablewidth{0pt}
\tablehead{
\colhead{Name (Epoch)} & \colhead{Ion} & \colhead{Comp.} &
\colhead{N (10$^{14}$ cm$^{-2}$)$^{a}$} &\colhead{}
}
\startdata
NGC 3516 (1) &C~IV &1  &8.0$\pm$0.5 &(8.4$\pm$0.5) \\
             &C~IV &2 &13.7$\pm$1.2 &(15.3$\pm$1.3) \\
             &C~IV &3  &0.4$\pm$0.1 & \\
             &C~IV &4  &0.5$\pm$0.1 & \\
NGC 3516 (2) &C~IV &1  &6.3$\pm$0.5 &(6.6$\pm$0.5) \\
             &C~IV &2 &11.0$\pm$2.4 &(15.4$\pm$3.3) \\
             &C~IV &3  &0.4$\pm$0.1 & \\
             &C~IV &4  &0.5$\pm$0.1 & \\
NGC 3783 (1) &N~V  &1  &$<$0.04     & \\
             &N~V  &2  &2.4$\pm$0.5 &(5.2$\pm$1.1)\\
NGC 3783 (2) &C~IV &1  &$<$0.05     & \\
             &C~IV &2  &$<$0.03     & \\
NGC 3783 (3) &C~IV &1  &$<$0.05     & \\
             &C~IV &2  &0.9$\pm$0.1 & \\
NGC 3783 (4) &C~IV &1  &1.2$\pm$0.2 & \\
             &C~IV &2  &0.9$\pm$0.3 & \\
NGC 5548 (1) &N~V  &1  &2.0$\pm$0.3 & \\
             &N~V  &2  &0.6$\pm$0.2 & \\
             &N~V  &3  &3.9$\pm$0.3 & \\
             &N~V  &4  &6.5$\pm$0.5 &(8.4$\pm$0.7) \\
             &N~V  &5  &1.1$\pm$0.2 & \\
NGC 5548 (2) &C~IV &1  &0.11$\pm$0.04 & \\
             &C~IV &2  &0.28$\pm$0.04 & \\
             &C~IV &3  &0.48$\pm$0.15 & \\
             &C~IV &4  &2.89$\pm$0.22 &(3.13$\pm$0.24) \\
             &C~IV &5  &0.41$\pm$0.18 & \\
\tablenotetext{a}{The first value assumes C$_{los}$ $=$ 1. The value in
parentheses was calculated using the C$_{los}$ value in Table 6.}
\enddata
\end{deluxetable}

\clearpage
\begin{figure}
\epsscale{0.9}
\plotone{fig1a.eps}
\\Fig.~1.
\end{figure}

\clearpage
\begin{figure}
\epsscale{0.9}
\plotone{fig1b.eps}
\\Fig.~1.--Continued
\end{figure}

\clearpage
\begin{figure}
\epsscale{0.9}
\plotone{fig2a.eps}
\\Fig.~2.
\end{figure}

\clearpage
\begin{figure}
\epsscale{0.9}
\plotone{fig2b.eps}
\\Fig.~2.--Continued
\end{figure}

\clearpage
\epsscale{0.9}
\begin{figure}
\plotone{fig2c.eps}
\\Fig.~2.--Continued
\end{figure}

\clearpage
\begin{figure}
\plotone{fig3.eps}
\\Fig.~3.
\end{figure}

\clearpage
\begin{figure}
\plotone{fig4.eps}
\\Fig.~4.
\end{figure}

\end{document}